\documentclass[twocolumn]{aastex7}

\received{August 19, 2025}
\revised{\today}

\submitjournal{AAS Journals}

\shorttitle{Hot Jupiters Were Once Cool Giant Planets}
\shortauthors{Schmidt \& Schlaufman}

\begin{document}

\title{Most Hot Jupiters Were Cool Giant Planets for More Than 1 Gyr}

\correspondingauthor{Stephen P.\ Schmidt}

\author[0000-0001-8510-7365]{Stephen P.\ Schmidt}
\altaffiliation{NSF Graduate Research Fellow}
\affiliation{William H.\ Miller III Department of Physics \& Astronomy,
Johns Hopkins University, 3400 N Charles Street, Baltimore, MD 21218, USA}
\email[show]{sschmi42@jhu.edu}

\author[0000-0001-5761-6779]{Kevin C.\ Schlaufman}
\affiliation{William H.\ Miller III Department of Physics \& Astronomy,
Johns Hopkins University, 3400 N Charles Street, Baltimore, MD 21218, USA}
\email{kschlaufman@jhu.edu}

\begin{abstract}

\noindent
The origin of hot Jupiters is the oldest problem in exoplanet
astrophysics.  Hot Jupiters formed in situ or via disk migration should
be in place just a few Myr after the formation of their host stars.
On the other hand, hot Jupiters formed via eccentricity excitation and
tidal damping as a result of planet--planet scattering or Kozai-Lidov
oscillations may take 1 Gyr or more to arrive at their observed locations.
We propose that the relative ages of hot Jupiters inside, near, and
outside the bias-corrected peak of the observed hot Jupiter period
distribution can be used to distinguish between these possibilities.
Though the lack of precise and accurate age inferences for isolated
hot Jupiter host stars makes this test difficult to implement,
comparisons between the Galactic velocity dispersions of the hot Jupiter
subpopulations enable this investigation.  To transform relative age
offsets into absolute age offsets, we calibrate the monotonically
increasing solar neighborhood age--velocity dispersion relation using
an all-sky sample of subgiants with precise ages and a metallicity
distribution matched to that of hot Jupiter hosts.  We find that the
inside-peak and near-peak subpopulations are older than the outside-peak
subpopulation, with the inside-peak subpopulation slightly older than
the near-peak subpopulation.  We conclude that at least 40\% but not more
than 70\% of the hot Jupiter population must have formed via a late-time,
peak-populating process like high-eccentricity migration that typically
occurs more than 1.5 Gyr after system formation.

\end{abstract}

\keywords{\uat{Exoplanet dynamics}{490} --- \uat{Exoplanet evolution}{491} ---
\uat{Exoplanet formation}{492} --- \uat{Exoplanet migration}{2205} ---
\uat{Exoplanets}{498} --- \uat{Stellar kinematics}{1608}}

\section{Introduction} \label{sec:intro}

Giant planets with orbital periods $P_{\text{orb}} < 10$ d (i.e., hot
Jupiters) form either in situ or at longer orbital periods.  In situ
formation requires massive or high dust-to-gas ratio protoplanetary
disks \citep[e.g.,][]{Bodenheimer00, Batygin16}.  In protoplanetary
disks more similar to the minimum-mass solar nebula however, giant planet
formation is expected take place at wider separations potentially beyond
a protoplanetary disk's water-ice line \citep[e.g.,][]{Pollack96,
Hubickyj05}.  These giant planets on wide-separation orbits only
become hot Jupiters if they experience inward migration.  This inward
migration can happen promptly as a result of interactions with a newly
formed planet's parent protoplanetary disk \citep[e.g.,][]{Ward97,
Lin96}.  Alternatively, this migration can occur after the era of
the protoplanetary disk as a result of planet--planet scattering
\citep[e.g.,][]{Rasio96, Weidenschilling96} or Kozai-Lidov oscillations
caused by a wide-separation stellar- or planetary-mass companion exciting
the proto-hot Jupiter's eccentricity \citep[e.g.,][]{Holman97, Mazeh97,
Kiseleva98, Wu03, Nagasawa08, Munoz16}.  Scattering events can occur
promptly after the dissipation of the parent protoplanetary disk or at
late times \citep[e.g.,][]{Beauge12, Wu11}, while Kozai-Lidov oscillations
always occur on secular timescales.  Tidal evolution also occurs on
secular timescales, with dissipation inside the planet circularizing
its orbit and aligning its rotation with the orbit's angular momentum.
In parallel, dissipation inside the star aligns its rotation with the
orbit's angular momentum and causes the orbit's semimajor axis to shrink
\citep[e.g.,][]{Rasio96b, Jackson08, Leconte10}.  These mechanisms make
differing predictions for the timescale of hot Jupiter formation as well
as the subsequent evolution of hot Jupiter systems.

The observed distribution of orbital parameters for hot Jupiter systems
have provided some insight into their formation pathways.  Studies of
their semimajor axis distribution \citep[e.g.,][]{Plavchan13, Nelson17},
their eccentricity distribution \citep[e.g.,][]{Petrovich15}, or both
\citep[e.g.,][]{Bonomo17} suggest that high-eccentricity migration can
explain most or all hot Jupiters within their selected samples.  Likewise,
studies of the hot Jupiter host star sky-projected obliquity distribution
indicate that high-eccentricity migration is a major (or even sole)
contributor to the hot Jupiter population \citep[e.g.,][]{Fabrycky09,
Morton11, Naoz12, Rice22}.

On the other hand, studies of the occurrence of wider-separation stellar-
or planetary-mass companions to hot Jupiters indicate that many hot
Jupiter systems lack perturbers capable of inducing high-eccentricity
migration \citep{Knutson14, Bryan16,Ngo16, Schlaufman16}.
Though some individual systems show evidence for high eccentricity
migration \citep[e.g.,][]{Naef01, Wu03, Gupta24}, other systems have
architectures impossible to reconcile with high-eccentricity migration
\citep[e.g.,][]{Becker15, Hord22}.  While these studies have provided
insight into the formation of hot Jupiters, inconsistencies still exist
and new approaches are needed to ultimately resolve hot Jupiter formation.

\begin{figure*}
    \centering
    \includegraphics[width=\linewidth]{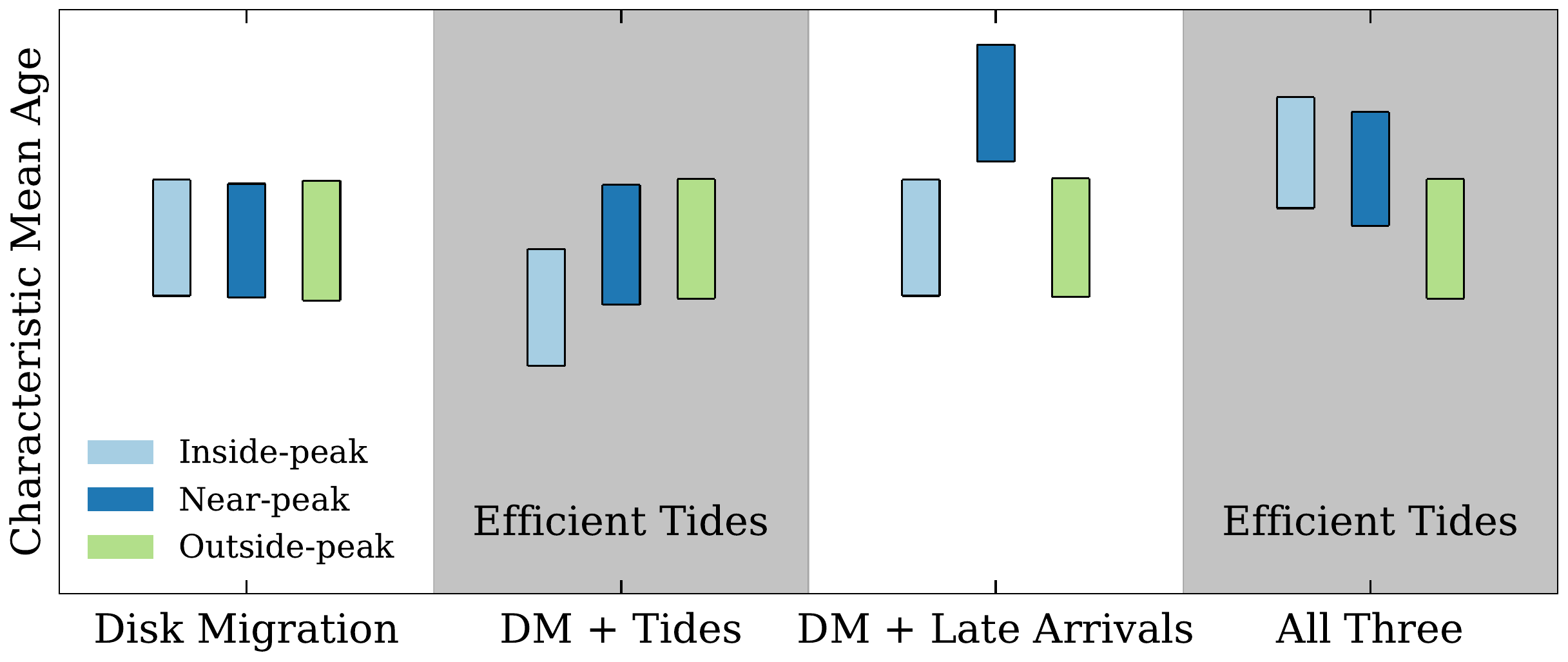}
    \caption{Cartoon of the characteristic mean ages expected for four
    different hot Jupiter formation scenarios: disk migration, disk
    migration plus tidal evolution, disk migration plus a late-time
    peak-populating mechanism, and disk migration plus a late-time
    peak-populating mechanism both affected by tidal evolution.  The light
    blue, dark blue, and light green rectangles correspond to the inside-,
    near-, and outside-peak subpopulations, while we indicate with gray
    shading the scenarios for which tidal dissipation is important.
    In the two scenarios on the left, hot Jupiters form almost entirely
    from an early-time process that results in an approximately uniform
    distribution in orbital period (e.g., disk migration).  In the two
    scenarios on the right, however, a large fraction of hot Jupiters
    form via a late-time process that results in a peaked orbital period
    distribution (e.g., high-eccentricity migration).  Each scenario
    results in different age orderings for the three subpopulations, so we
    argue that characteristic mean age measurements should differentiate
    between them.}
    \label{fig:scenarios}
\end{figure*}

New techniques that provide access to the time evolution of hot Jupiter
systems have recently been used to constrain their eventual fates.
\citet{Hamer19} used a sample of solar neighborhood hot Jupiters
available at that time to show that the Galactic velocity dispersion
of their host stars was much colder than matched control samples of
field stars, suggesting that the hot Jupiter population is relatively
young as a consequence of the tidal destruction of the oldest and
smallest-separation systems.  This result was independently confirmed
by \citet{Miyazaki23} and \citet{Chen23}.  \citet{Hamer20} subsequently
obtained similar results for Kepler-discovered hot Jupiters.

More recent population age calculations have moved beyond studying the
fates of hot Jupiters towards the time evolution of planetary systems.
\citet{Hamer22} showed that misaligned hot Jupiter systems have a
warmer velocity dispersion than aligned systems.  They interpreted this
observation as evidence that hot Jupiters that formed via a process that
increases host star obliquity arrived later than hot Jupiters formed
via a process that favors low host star obliquities.  This approach
becomes even more powerful if the age--velocity dispersion relation
can be calibrated to provide absolute population age as a function
of Galactic velocity dispersion.  \citet{Schmidt24} calibrated the
age--velocity dispersion relation in the Kepler field and showed that
ultra-short-period planets arrived at their observed locations after
billions of years of tidal migration.

A calibrated solar neighborhood age--velocity dispersion relation could
be used to study the formation time evolution of hot Jupiter systems
more precisely than previously possible.  A hot Jupiter population formed
primarily via in situ formation or disk migration would include systems
as young as a few Myr.  On the other hand, planet--planet scattering need
not take place immediately after the dissipation of planetary systems'
parent protoplanetary disks.  Similarly, Kozai-Lidov oscillations occur
only on secular timescales.  As a result, a hot Jupiter population
formed primarily via high-eccentricity migration could have relatively
fewer younger systems and an older average age than a population formed
primarily in situ or via disk migration.  Subsequent tidal dissipation
can also affect the observed mean ages of the smallest-separation hot
Jupiter systems.

To clarify the relative importance of these formation mechanisms for the
overall hot Jupiter population, we propose to compare the characteristic
mean ages of different hot Jupiter subpopulations as illustrated in
Figure \ref{fig:scenarios}.  We first subdivide the overall hot Jupiter
population into three subpopulations based on their proximity to the
peak of the orbital period distribution after accounting for observation
biases: (1) an inside-peak subpopulation, (2) a near-peak subpopulation,
and (3) an outside-peak subpopulation.  If an early time mechanism like
disk migration or in situ formation is responsible for most hot Jupiters,
then all three subpopulations would have similar characteristic mean ages.
If tidal evolution is efficient, then the inside-peak subpopulation will
lose its oldest members and will therefore have a younger characteristic
mean age; the other two subpopulations would be much less affected.
If superimposed on an early-time population is a larger population of
hot Jupiters formed via a late-time mechanism like high-eccentricity
migration, then the preference for high-eccentricity migration to populate
the orbital period peak would result in an older near-peak subpopulation.
If both early- and late-time mechanisms operate and tidal evolution is
efficient, then the inside-peak subpopulation would be replenished by
the older near-peak subpopulation, thereby increasing its characteristic
mean age.  At the same time, because the outside-peak subpopulation
would have mostly formed via the early-time mechanism and be largely
unaffected by tides, it would have the youngest characteristic mean age.

In this article we execute the test described in the preceding paragraph.
We assemble our solar neighborhood subgiant sample and hot Jupiter
subpopulations in Section \ref{sec:sample}.  We then calibrate the
age--velocity dispersion relation, calculate the subpopulations'
velocity dispersions, and obtain characteristic mean ages for each
subpopulation in Section \ref{sec:methods}.  We discuss the implications
of our analyses in Section \ref{sec:disc} and summarize our findings in
Section \ref{sec:summ}.

\section{Data} \label{sec:sample}

New searches for hot Jupiters at longer orbital periods have enabled
our investigation into the relative importance of hot Jupiter formation
mechanisms.  Most of these planets have been discovered by NASA's
Transiting Exoplanet Survey Satellite \citep[TESS;][]{Ricker14}, as its
coverage of nearly the whole sky has enabled a magnitude-limited search
for transiting hot Jupiters out to $P\approx10$ d and $T\approx12$
\citep[e.g.,][]{Yee22, Yee23, Yee25, Schulte24}.  These new
discoveries along with Gaia Data Release (DR) 3 astrometry and radial
velocities\footnote{For the details of Gaia DR3 and its data processing,
see \citet{Gaia_mission2016, GaiaEDR3, GaiaDR3Frame, GaiaDR3Summary},
\citet{GaiaEDR3Validation}, \citet{lin21a, lin21b}, \citet{GaiaEDR3XMatch,
GaiaDR3XMatch}, \citet{EDR3photometry}, \citet{row21}, \citet{tor21}, and
\citet{GaiaDR3Validation}.} provide the data necessary for our experiment.

We select from the NASA Exoplanet Archive's \texttt{pscomppars} table
\citep{ExoplanetArchive, Christiansen25, PSCompPars} as of July 22,
2025 a sample of 503 hot Jupiters that meet the following criteria:
P$_{\text{orb}} < 10$ d and $0.1 < M_{p} < 10$ M$_{\text{Jup}}$.
We use the hot Jupiter orbital period and mass lower limit advocated
by \citet{Wright12} and the giant planet mass upper limit proposed by
\citet{Schlaufman18}.  Due to Doppler surveyors' historical avoidance
of stars with observable emission in the cores of the Ca H and K
lines, Doppler surveys have an intrinsic bias against young stars
\citep{Marcy05}.  We select only transiting systems to avoid this age
bias in our analysis sample.  Our approach also removes objects with only
a minimum mass inference that may exceed our planet mass upper limit.
Because the solar neighborhood age--velocity dispersion relation may
not be the same as the age--velocity dispersion relation in the Kepler
field \citep{Schmidt24}, we exclude systems discovered by Kepler as well.

To calculate the Galactic velocity dispersions necessary to conduct
our experiment, we require Gaia astrometry.  We require Gaia DR3
\texttt{source\_id} identifiers for every system in our analysis sample
by first using the Gaia DR2 \texttt{source\_id} identifier provided by
the NASA Exoplanet Archive to find the DR3 \texttt{source\_id} using
the \texttt{dr2\_neighbourhood} table.  If a NASA Exoplanet Archive
entry lacks a Gaia DR2 \texttt{source\_id} but has a TESS Input Catalog
\citep[TIC; ][]{TIC2019} ID, then we use the TIC ID to obtain a Gaia
DR2 \texttt{source\_id} and then follow the same procedure.  If a NASA
Exoplanet Archive entry lacks both a Gaia DR2 \texttt{source\_id} and
a TIC ID, then we use \texttt{astroquery} \citep{astroquery} to query
Simbad for the corresponding Gaia DR3 \texttt{source\_id} corresponding
to the system \texttt{hostname}.

We use the Gaia DR3 \texttt{source\_id} identifiers and TOPCAT
\citep{TOPCAT} to select equatorial positions, proper motions,
parallaxes, and all covariances between these parameters for our
analysis sample.  We additionally collect radial velocities and
associated uncertainties \citep{DR3_velocities}.  To ensure high-quality
Galactic $UVW$ velocity measurements, we follow \citet{Hamer24} and
\citet{Schmidt24} in requiring that \texttt{parallax\_over\_error} $>10$,
\texttt{rv\_nb\_transits} $>10$, and \texttt{rv\_expected\_sig\_to\_noise}
$>5$.  To avoid unresolved binaries, we also require \texttt{ruwe} $<
1.4$ \citep{Ziegler2020}.  We also collect the astrometric solution data
necessary to correct for parallax zero-point offsets.

We supplement the radial velocities for our hot Jupiter sample by
obtaining Apache Point Observatory Galactic Evolution Experiment
\citep[APOGEE;][]{APOGEE} heliocentric radial velocities and their
associated uncertainties.  This data was derived from spectra that were
gathered during the third and fourth phases of the Sloan Digital Sky
Survey \citep[SDSS;][]{SDSSIII,SDSSIV} as part of APOGEE.  These spectra
were collected with the APOGEE spectrographs \citep{Zas13, Zas17,
Wil19, Bea21, San21} on the New Mexico State University 1-m Telescope
\citep{Hol10} and the Sloan Foundation 2.5-m Telescope \citep{SDSS2.5m}.
As part of SDSS DR 17 \citep{Abdurrouf22}, these spectra were reduced
and analyzed with the APOGEE Stellar Parameter and Chemical Abundance
Pipeline \citep[ASPCAP; ][]{All06, Hol15, Nid15, ASPCAP} using an $H$-band
line list, MARCS model atmospheres, and model-fitting tools optimized for
the APOGEE effort \citep{Alv98, Gus08, Hub11, Ple12, Smi13, Smi21, Cun15,
She15, Jonsson20}.  We prefer these radial velocities over Gaia DR3 radial
velocities in situations where both are available.  It is valid to include
both APOGEE and Gaia DR3 radial velocities due to their consistent zero
points \citep[e.g.,][]{Schmidt24}.  We list the host Jupiter hosts in
our analysis sample with their Gaia DR3 \texttt{source\_id} identifiers,
radial velocities, and radial velocity sources in Table \ref{tab:samples}.

\begin{deluxetable*}{lccccccc}
    \centering
    \tabletypesize{\scriptsize}
    \tablewidth{0pt}
    \tablecaption{Hot Jupiter System Samples\label{tab:samples}}
    \tablehead{
    \colhead{System Name} &
    \colhead{Gaia DR3 \texttt{source\_id}} &
    \colhead{RV} &
    \colhead{RV Source} &
    \colhead{Subpopulation}\\
    & & (km s$^{-1}$) &}
    \startdata
    \hline
    K2-77 & 37619725922094336 & $7.90 \pm 0.03$ & APOGEE & Outside-peak \\
    HD 285507 & 45159901786885632 & $38.02 \pm 0.17$ & Gaia & Outside-peak \\
    K2-87 & 47908509057301120 & $16.54 \pm 0.02$ & APOGEE & Outside-peak \\
    V1298 Tau & 51886335968692480 & $15.62 \pm 0.06$ & APOGEE & Outside-peak \\
    TOI-5344 & 52359538285081728 & $45.00 \pm 2.63$ & Gaia & Near-peak \\
    K2-30 & 61607255708760960 & $36.18 \pm 0.04$ & APOGEE & Near-peak \\
    HAT-P-25 & 111322601672419712 & $-13.72 \pm 1.06$ & Gaia & Near-peak \\
    HAT-P-52 & 128623485853170432 & $60.92 \pm 2.93$ & Gaia & Inside-peak \\
    V830 Tau & 147831571737487488 & $18.00 \pm 0.05$ & APOGEE & Outside-peak \\
    K2-29 & 150054788545735424 & $32.77 \pm 0.48$ & Gaia & Inside-peak \\
    \enddata
    \tablecomments{This table sorted by Gaia DR3 \texttt{source\_id}
    and is published in its entirety in machine-readable format.}
\end{deluxetable*}

Hot Jupiters tend to orbit metal-rich thin disk stars
\citep[e.g.,][]{Santos04, Fischer05}.  To ensure that our analysis sample
is not contaminated by the thick disk, we calculate the maximum height
above the mid plane $z_{\text{max}}$ for each system in our analysis
sample.  We integrate each star's Galactic orbit from its position,
parallax, proper motions, and radial velocity using the \texttt{gala}
Python package \citep{gala}.  Regardless of our assumption about
the Galaxy's gravitational potential used to integrate these orbits
\citep[e.g., variations on those presented by][]{Bovy15}, we find that no
more than 10 hot Jupiter systems have a $z_{\text{max}}$ greater than 1
kpc, a potential indicator for thick disk membership.\footnote{We note
that one system, the metal-poor WASP-183 system, has a particularly
dynamically warm orbit.} We therefore conclude that no more than 2\%
of our sample is in the high-velocity tail of the solar neighborhood
velocity distribution.

To calibrate the solar neighborhood age--velocity dispersion relation,
we follow the same procedure described above for the sample of subgiants
with accurate, precise, and physically self-consistent age inferences
presented in \citet{Nataf24}.  As small differences in mass result in
large differences in observed luminosity for stars transitioning from
core to shell hydrogen fusion, subgiants represent the optimal stage
of stellar evolution for age inference.  \citet{Nataf24} used MESA
Isochrones and Stellar Tracks \citep[MIST;][]{pax11, pax13, pax18, pax19,
jer23, dot16, cho16} models and the \texttt{isochrones} Python package
to infer fundamental and photospheric stellar parameters for 401,821
stars by fitting the MIST models to precise zero point-corrected Gaia DR3
parallaxes, multi-wavelength photometry, and extinction measurements from
3D reddening maps.  The \citet{Nataf24} catalog reports typical random
uncertainties on age of $\sigma_\tau\approx8\%$ and has been extensively
validated with other catalogs of subgiant parameters.  To enable a
consistent comparison between the calibrated age--velocity dispersion
relation and our hot Jupiter subpopulations' velocity dispersions,
we next remove all subgiants from this sample with $z_{\text{max}} >
0.4$ kpc.  This results in 88366 subgiants that pass all data quality
cuts and have kinematics suggestive of thin disk membership.

\section{Analysis} \label{sec:methods}

To execute our experiment, we must divide our sample into inside-peak,
near-peak, and outside-peak hot Jupiter subpopulations.  While this
subdivision could be made in several observable or inferable properties
like orbital period $P_{\text{orb}}$, semimajor axis $a$, or scaled
semimajor axis $a/R_{\ast}$, we favor and therefore use $P_{\text{orb}}$
to subdivide the hot Jupiter sample because, unlike $a$ and $a/R_{\ast}$,
$P_{\text{orb}}$ is directly observable from the light curves of all
transiting exoplanet systems.  Like all samples of transit-discovered
exoplanets, our transiting hot Jupiter sample is biased as a consequence
of the probability of transit:
\begin{equation}
    P(\text{transit}) \propto \left(\frac{a}{R_\ast}\right)^{-1},
\end{equation}
where $a/R_\ast$ is the scaled semimajor axis of the star-planet system.
To address this bias, we therefore calculate the debiased median
orbital period using the reported scaled semimajor axes from the NASA
Exoplanet Archive as weights.  We define the resulting orbital period
$P_{\text{orb}} = $ 3.92 d as the debiased peak of our analysis sample.
To divide our analysis sample into the subpopulations described above,
we first create the near-peak subpopulation by selecting one third
of the analysis sample with orbital periods closest to the debiased
peak.  The resulting upper and lower orbital period limits for this
subpopulation are at 3.259 and 4.545 d.  Our inferred debiased orbital
period peak is consistent with the hot Jupiter occurrence peak observed
by \citet{Santerne16} among confirmed Kepler-discovered giant planets.
We next place all hot Jupiters with $P_{\text{orb}} < 3.259$ d into
the inside-peak subpopulation and likewise place all hot Jupiters with
$P_{\text{orb}} > 4.545$ into the outside-peak subpopulation.  We list
the subpopulation membership of the systems in our analysis sample in
Table \ref{tab:samples}.

\begin{figure*}
    \centering
    \includegraphics[width=\linewidth]{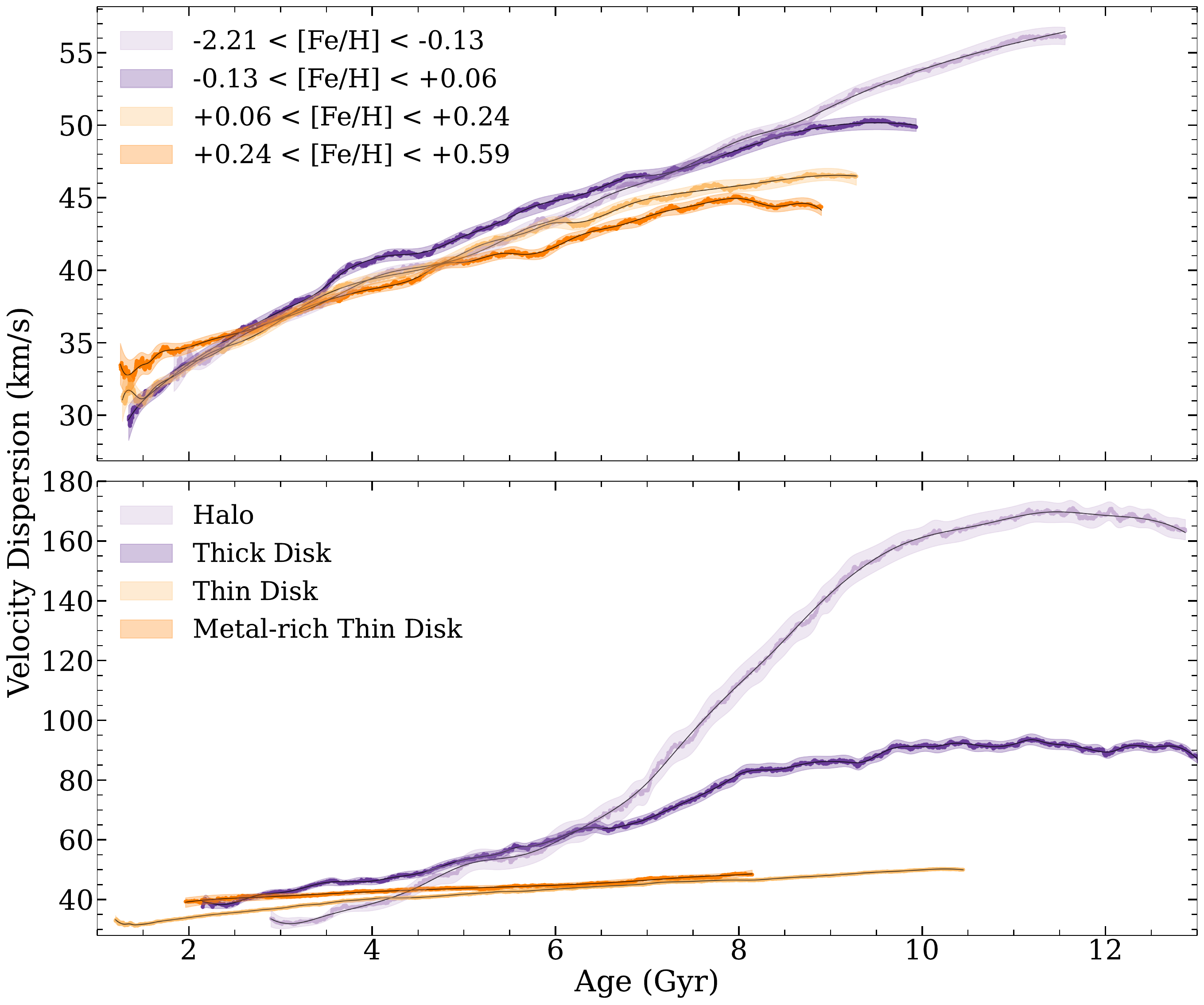}
    \caption{Solar neighborhood age--velocity dispersion relations as a
    function of metallicity. We subdivide the subgiant sample presented
    in \citet{Nataf24} into four equal-size metallicity bins. We then
    order each subsample in age and calculate velocity dispersion in
    consecutive windows of 3000 stars.  We plot the result of these
    calculations as overlapping colored points: light purple for the
    first quartile, dark purple for the second quartile, light orange
    for the third quartile, and dark orange for the fourth quartile. We
    plot as solid colored lines a smoothing spline of the respective
    data and as transparent polygons the 16th/84th interquantile ranges
    of the subsample's velocity dispersion distribution suggested by
    bootstrap resampling. Though they exhibit noticeable differences
    in velocity dispersion at older ages, there is no significant
    metallicity dependence on velocity dispersion between 2 and 4 Gyr,
    the characteristic mean ages of our hot Jupiter subpopulations.}
    \label{fig:metalcomp}
\end{figure*}

It is well known that the occurrence of hot Jupiters increases strongly
with host star metallicity \citep{Santos04, Fischer05}.  For that
reason, a possible dependence of the solar neighborhood age--velocity
dispersion relation on metallicity could bias the characteristic
mean ages we infer for our analysis sample.  To investigate this
possibility, we divide our subgiant sample into four equal-size bins in
metallicity: $-2.21<\text{[Fe/H]} <-0.13$, $-0.13<\text{[Fe/H]} <+0.06$,
$+0.06<\text{[Fe/H]} <+0.24$, and $+0.24<\text{[Fe/H]} <+0.59$.  Using a
window size of 3000, we follow the methodology of \citet{Schmidt24} to
construct age--velocity dispersion relations for each metallicity bin.
We show the results of these calculations in Figure \ref{fig:metalcomp}.
In the interval 2 Gyr $< \tau<$ 4 Gyr, appropriate for our subsequent
analyses, the age--velocity dispersion relation is independent of
metallicity.  Though we will restrict our subgiant sample to a similar
metallicity range to that spanned by our hot Jupiter sample for the
calibration of our age--velocity dispersion relation, the effect of
metallicity on this calibration is negligible.

We use the \texttt{pyia} Python package
\citep{adrian_price_whelan_2018_1228136} to convert equatorial
coordinates, proper motions, parallaxes, and radial velocities
into Galactic $UVW$ space velocities.  As has been the case with
our previous kinematic analyses, the individual radial velocity
measurement uncertainties are an order of magnitude smaller than our
velocity dispersion measurements, so our analyses are not limited by
radial velocity precision.  Prior to calculation of velocities, we
use the \texttt{gaiadr3\_zeropoint} Python package to correct each
star's parallax measurement \citep{lin21a}.  We use a Monte Carlo
simulation in which \texttt{pyia} randomly samples from the Gaia
DR3 five-parameter astrometric solution respecting each solution's
covariance.  It also independently randomly samples radial velocities
from a normal distribution with mean and variance as reported in each
radial velocity source.  We sample 100 realizations from each star's
astrometric uncertainty distributions in position, proper motion,
parallax, and radial velocity using \texttt{pyia}.  We then calculate
the velocity dispersion $\sigma$ of each sample
\begin{equation} \label{eq:vd}
    \sigma = \frac{1}{N} \sum \limits_{i=1}^{N} \left[(U_i - \bar{U})^2 + (V_i - \bar{V})^2 + (W_i - \bar{W})^2 \right]^{1/2},
\end{equation}
to construct an ensemble of samples of the entire populations' velocity
dispersions.  As was also the case in our previous velocity dispersion
analyses \citep{Hamer19, Hamer20, Hamer22, Hamer24, Schmidt24}, $\bar{U}$,
$\bar{V}$, and $\bar{W}$ are the median $U$, $V$, and $W$ velocities of
the sample under consideration.

\begin{figure*}
    \centering
    \includegraphics[width=\linewidth]{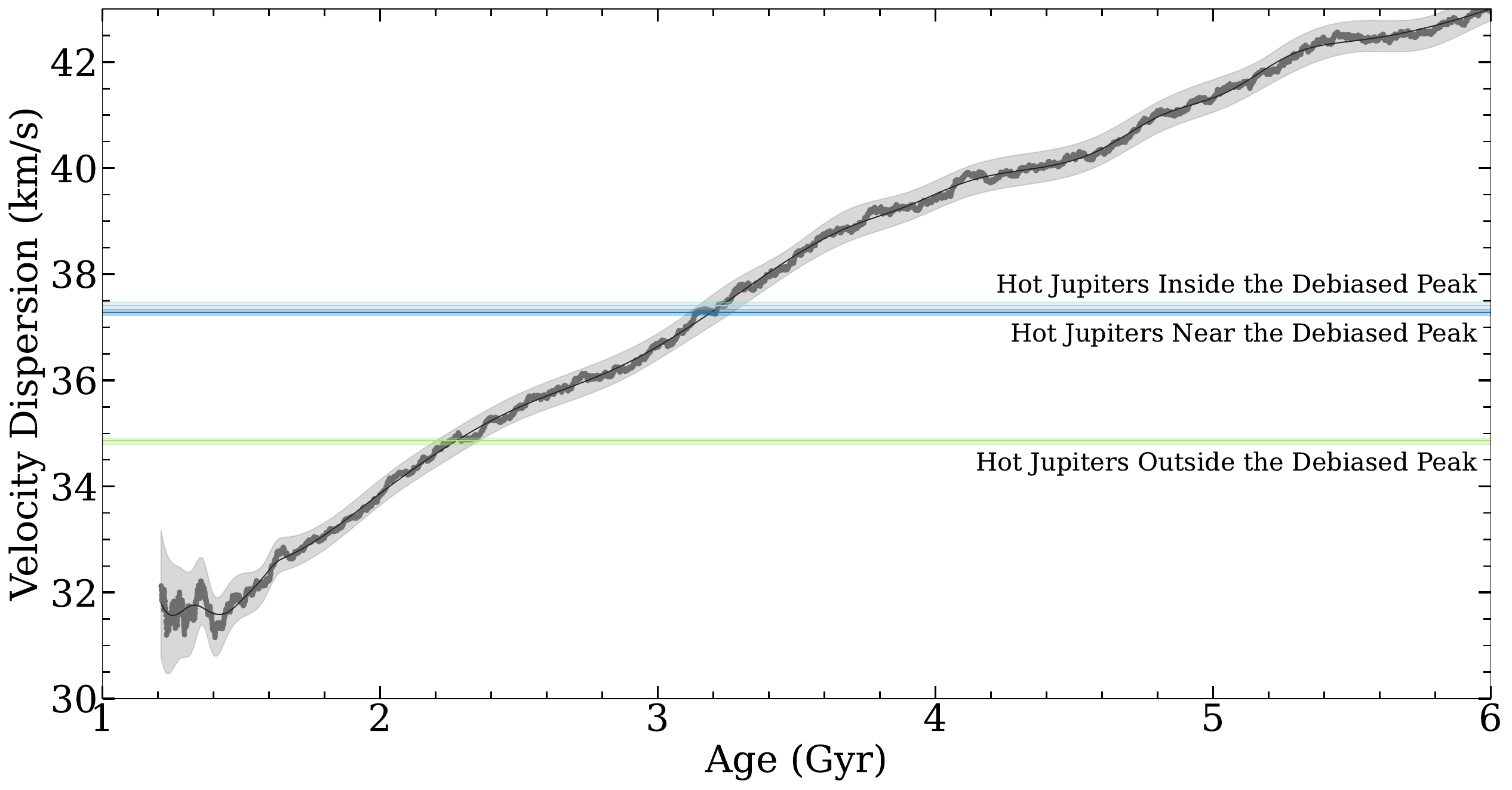}
    \caption{Solar neighborhood age--velocity dispersion relation
    for the metallicity interval $-0.06 < \text{[Fe/H]} < +0.36$
    appropriate for hot Jupiter hosts as explained in Section
    \ref{sec:methods}.  We follow the same procedure we use to generate
    Figure \ref{fig:metalcomp} and plot as the black line the resulting
    smoothed age--velocity dispersion relation and as the gray polygon
    its 16th/84th interquantile range.  We calculate the debiased
    median period of the hot Jupiter period distribution by weighting
    hot Jupiter systems' orbital periods by their scaled semimajor
    axes $a/R_\ast$ to account for the observational biases present
    in the transit technique that has discovered most hot Jupiters.
    The near-peak subpopulation comprises one third of the overall
    sample with orbital periods closest to this debiased median period.
    We then create two additional subpopulations: one inside the debiased
    peak with orbital periods shorter than the near-peak subpopulation
    and one outside the debiased peak with orbital periods longer than
    the near-peak subpopulation.  We plot as light blue, dark blue,
    and light green horizontal lines the velocity dispersions of
    the inside-, near-, and outside-peak hot Jupiter subpopulations.
    The inside-peak and near-peak subpopulations have characteristic
    mean ages $\tau \approx 3.2$ Gyr, with the inside-peak subpopulation
    slightly older than the near-peak subpopulation.  On the other hand,
    the outside-peak subpopulation has a characteristic mean age $\tau
    \approx 2.3$ Gyr.  The distinctly younger age of the outside-peak
    subpopulation relative to the inside- and near-peak subpopulations
    supports the fourth scenario in Figure \ref{fig:scenarios}.
    In that scenario, the combination of (1) an early-time formation
    channel that produces a uniform period distribution, (2) a late-time
    formation channel that produces a peaked period distribution, and
    (3) subsequent tidal evolution have all contributed to the formation
    and evolution of the hot Jupiter population.}
    \label{fig:thirdscomp}
\end{figure*}

We use the same procedure to calculate $UVW$ velocities for our sample
of subgiants.  According to the NASA Exoplanet Archive the interval
between the 16th and 84th quantiles of our hot Jupiter sample's host
stars metallicity distribution is $(-0.06, 0.36)$.  To generate our
calibrated age--velocity dispersion relation appropriate for comparison
to our analysis sample, we therefore select subgiants with 1-$\sigma$
uncertainties in [Fe/H] that overlap with this interval.  This reduces
our sample to 47390 subgiants.  We next sort this metallicity-matched
subgiant sample by age and calculate the mean age and velocity dispersion
in a moving window sample of 5000 subgiants. We choose a window of
5000 to balance time resolution and velocity dispersion precision.
This window is much larger than the window we use in \citet{Schmidt24},
as our metallicity-matched subgiant sample is more than an order
of magnitude larger.  We then advance the window by one star and
then repeat this procedure until we cover the entire age range of our
subgiant sample.  For each 5000-star window, we sample with replacement
within the window 150 bootstrap resamples of 5000 stars and calculate the
velocity dispersion of each sample.  We report in Table \ref{tab:ageVD}
the 16th, 50th, and 84th quantiles of the resulting velocity dispersion
distributions.  In parallel, we calculate the average age of the 5000
stars in each window.  At the youngest ages, we use a reduced window
size to maximize age resolution.  We follow the same procedure described
above, but we start each window with the youngest star and sequentially
increase the window width from 150 to 5000 as the window advances.
In contrast to previous studies of the relationship between age and
velocity dispersion in the solar neighborhood \citep[e.g.,][]{Aumer09,
Chen21}, our nonparametric approach does not assume or fit any functional
form to the relationship.  We use univariate smoothing splines to smooth
the curves connecting the 16th, 50th, and 84th quantiles of the velocity
dispersion distributions in each window, and we plot this smoothed
age--velocity dispersion relation in Figure \ref{fig:thirdscomp}.

\begin{deluxetable*}{cccc}
    \centering
    \tabletypesize{\scriptsize}
    \tablewidth{0pt}
    \tablecaption{Solar Neighborhood Age--Velocity Dispersion Relation
    for Hot Jupiter Hosts\label{tab:ageVD}}
    \tablehead{
    \colhead{Window Average Age} &
    \colhead{Lower Uncertainty} &
    \colhead{Velocity Dispersion} &
    \colhead{Upper Uncertainty} \\
    (Gyr) & (km s$^{-1}$) & (km s$^{-1}$) & (km s$^{-1}$)}
    \startdata
    1.21211	& 1.28785	& 32.12854	& 1.23260\\
    1.21268	& 1.33804	& 32.11174	& 1.29713\\
    1.21324	& 1.33564	& 32.05038	& 1.40146\\
    1.21379	& 1.26139	& 31.95224	& 1.20396\\
    1.21435	& 1.02253	& 31.86713	& 1.00291\\
    1.21490	& 1.17319	& 31.92756	& 1.26564\\
    1.21544	& 1.15702	& 31.98542	& 1.23763\\
    1.21597	& 1.33164	& 31.83756	& 1.40988\\
    1.21651	& 1.05254	& 31.83700	& 1.07099\\
    1.21704	& 1.25875	& 31.84092	& 1.38797\\
    \enddata
    \tablecomments{This table is ordered by average age in ascending order
    and is published in its entirety in machine-readable format. This
    age--velocity dispersion relation uses subgiants in a metallicity
    range restricted to reflect that of hot Jupiter hosts. Lower
    uncertainty refers to the difference between the 50th and 16th
    quantiles, and the upper uncertainty refers to the difference between
    the 84th and 50th quantiles.}
\end{deluxetable*}

\begin{figure*}
    \centering
    \includegraphics[width=\linewidth]{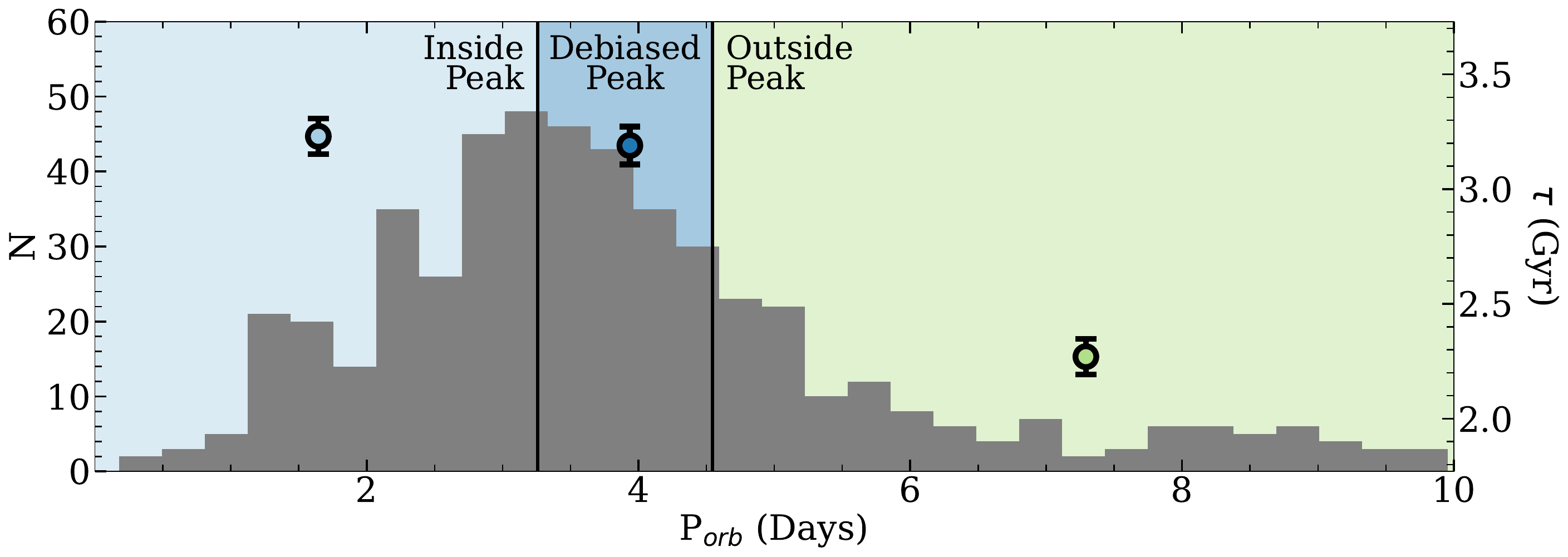}
    \caption{Illustration of the three hot Jupiter subpopulations and
    their characteristic mean ages. We plot as the gray histogram the
    orbital period distribution of our solar neighborhood transiting hot
    Jupiter sample.  We mark with vertical black lines the boundaries of
    the three subpopulations we have identified and shade the inside-,
    near-, and outside-peak subpopulations in light blue, dark blue,
    and light green.  We plot the characteristic mean age ranges as
    the black error bars in each region.  We find that the inside-peak
    subpopulation is older, but still statistically consistent with,
    the characteristic mean age of the near-peak subpopulation.
    In contrast, the outside-peak subpopulation is younger than both
    shorter-period subpopulations.  We argue that this ordering in
    age plus the statistically younger characteristic mean age of
    the outside-peak subpopulation is best explained by a late-time,
    peak-populating formation mechanism that is responsible for placing
    most hot Jupiters near and inside the debiased orbital period peak.
    Subsequent tidal evolution would then move some of these hot Jupiters
    to shorter orbital periods over billions of years.}
    \label{fig:perioddist}
\end{figure*}

We next use this age--velocity dispersion relation to calculate
velocity dispersion-based characteristic mean ages for our hot
Jupiter subpopulations represented in Table \ref{tab:samples}.
As in \citet{Schmidt24}, we infer lower and upper limits for these
characteristic mean ages by identifying the range over which our inferred
planet population mean velocity dispersions overlap with the 1-$\sigma$
range of the metallicity-constrained solar neighborhood age--velocity
dispersion relation.  We argue that this approach to age inferences is
closest to the data itself.  Conversely, other approaches that have
assumed parametric relationships between age and velocity dispersion
must account for random uncertainties in both their velocity dispersion
measurements and the parameters of their parametric models like power law
exponents \citep[e.g.,][]{Chen21}.  We find that the characteristic mean
age ranges for our inside-peak, near-peak, and outside-peak subpopulations
are $ 3.16~\text{Gyr}< \tau < 3.30$ Gyr, $3.11~\text{Gyr}< \tau < 3.27$
Gyr, and $2.20~\text{Gyr}< \tau < 2.36$ Gyr. We show these ranges in
Figure \ref{fig:perioddist}.

We find that the characteristic mean age of the outside-peak
subpopulation is about 750 Myr younger than the near-peak and inside-peak
subpopulations.  In addition, the inside-peak subpopulation is slightly
older than the near-peak subpopulation.  This ordering supports the
fourth scenario shown in Figure \ref{fig:scenarios} and indicates that the
formation and evolution of hot Jupiters can be attributed to meaningful
contributions from an early-time, uniformly-distributing mechanism,
a late-time, peak-populating mechanism, and tidal evolution.

Some potential observational biases could produce our observed age offset.
As a system's transit probability is approximately the inverse of its
scaled semimajor axis, hot Jupiters around more massive and therefore
younger and larger host stars would be more likely to transit.  However,
the same is also true for hot Jupiters around older stars due to the
radius expansion stars experience as they evolve on the main sequence.

To assess the age bias incurred by these competing effects, we construct
a synthetic solar neighborhood sample of hot Jupiter host stars starting
from Gaia DR3's \texttt{gaia\_universe\_model} data product.  We select
dwarf stars with $\log g > 3.75$ and $G < 12$ as an approximate limit for
a ``complete'' magnitude-limited hot Jupiter population \citep{Yee21}.
To match the metallicity of this synthetic sample to our analysis sample,
we also restrict it to $-0.06 < \text{[Fe/H]} < 0.36$.  Because we are
only concerned with age differences resulting from the bias described
above and not the overall occurrence of hot Jupiters, to maximize the
precision of our bias quantification it is acceptable to assume that
all stars in our synthetic sample host hot Jupiters.  We then perform a
Gaussian kernel density estimation on our debiased hot Jupiter period
distribution to generate orbital periods for each synthetic system.
We next assume an isotropic inclination distribution and select
transiting systems following Equation (1) of \citet{Gilliland00}.
Dividing each realization's entire population into inside-peak, near-peak,
and outside-peak subpopulations, we calculate the median age of each
subpopulation.  Finally, we create an ensemble of age differences between
the near-peak and outside-peak subpopulations to assess a potential age
bias.  We find that the 1-$\sigma$ range of the age offset distribution
is consistent with zero whether or not we apply a metallicity cut.
The same is true when we relax the apparent $G$ magnitude limit to $G<14$
regardless of metallicity.  Likewise, there is no change if we exclude
stars close to the ecliptic plane that have been less intensively studied
by TESS than stars at the ecliptic poles.  As a result, we can conclude
that any age differences we observe are caused by the time evolution of
hot Jupiter systems.  We note that there is no relationship between age
and metallicity in the thin disk of the Milky Way \citep[e.g.,][]{Xiang22,
Nataf24}, so there is no need to consider biases that might result from
a relationship between age and metallicity in our sample.

\section{Discussion} \label{sec:disc}

In our solar neighborhood hot Jupiter analysis sample, we find that the
outside-peak subpopulation has a characteristic mean age in the range
$2.20~\text{Gyr} \lesssim \tau \lesssim 2.36$ Gyr.  On the other hand,
we find that the near- and inside-peak subpopulations have characteristic
mean ages in the ranges $3.11~\text{Gyr} \lesssim \tau \lesssim 3.27$
Gyr and $3.16~\text{Gyr} \lesssim \tau \lesssim 3.30$ Gyr.  We emphasize
that the characteristic mean ages we quote are statistical measurements
of population mean ages, and that individual systems may be older or
younger than their subpopulation's characteristic mean age.

\begin{figure*}
    \centering
    \includegraphics[width=\linewidth]{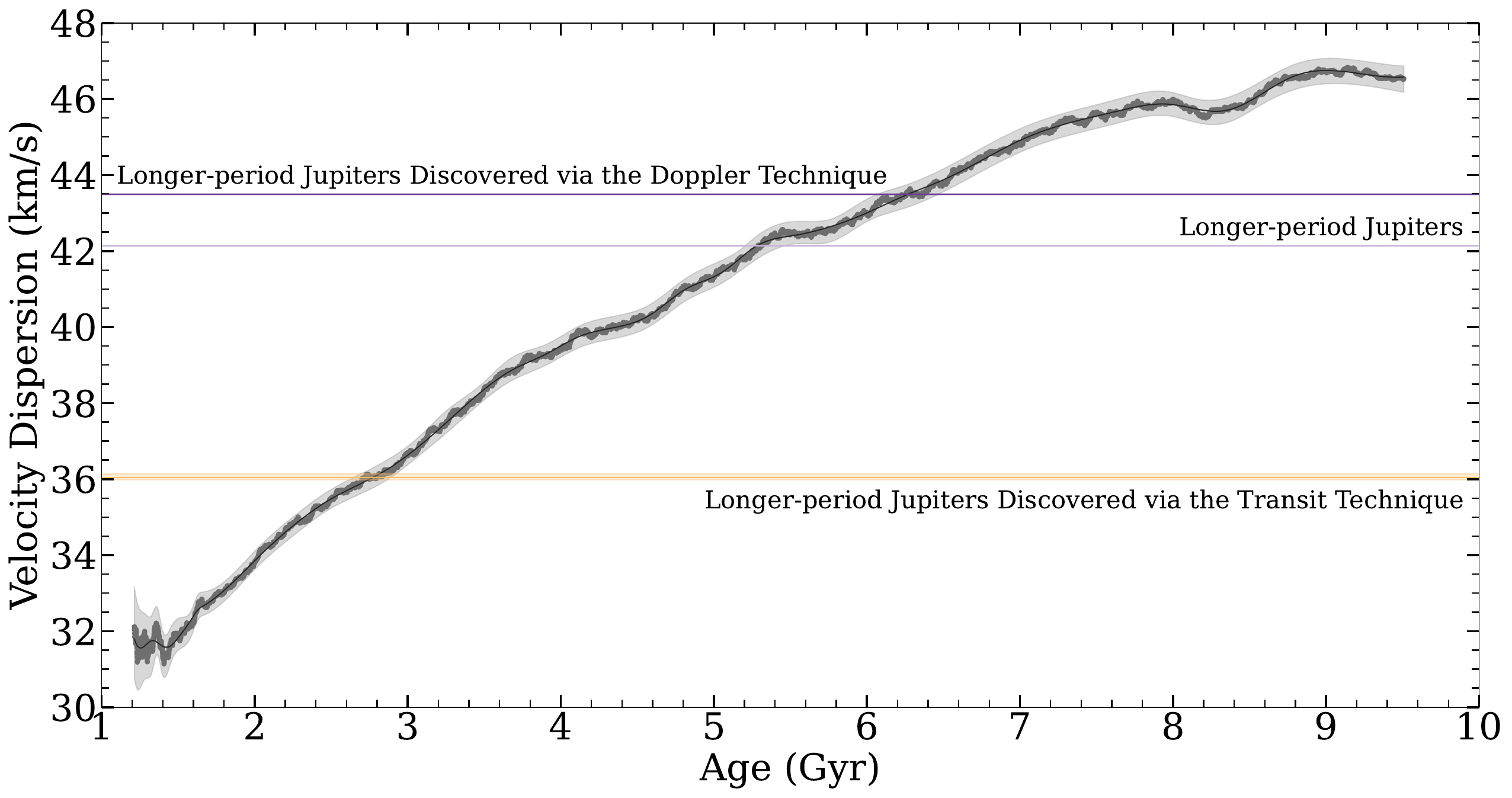}
    \caption{The same age--velocity dispersion relation as in Figure
    \ref{fig:thirdscomp}.  To emphasize the effect of observational
    biases on the characteristic mean ages of exoplanet systems, we
    plot as horizontal lines the velocity dispersions of all known
    giant planet systems with orbital periods $P>10$ d (light purple),
    the subset of these systems discovered via the Doppler technique
    (dark purple), and the subset of these systems discovered via
    the transit technique (orange).  Because by itself the Doppler
    technique is only sensitive to long-period giant planets orbiting
    quiescent stars, Doppler surveyors have usually avoided stars with
    chromospheric emission indicative of youth.  This bias against young
    stars in the population of Doppler-discovered exoplanet systems
    makes comparing the mean ages of the Doppler- and transit-discovered
    exoplanet populations challenging.  We caution against combining
    populations of planets detected via more than one technique for
    system age-related analyses without first accounting for biases
    incurred by each detection method.}
    \label{fig:methodscomp}
\end{figure*}

We highlight the importance of analyzing transit-discovered systems
separately from Doppler-discovered systems because of the selection
effects that shape both samples.  In particular, Doppler surveyors tend
to avoid young stars because young stars have large radial velocity
jitters that make detection and characterization of small Doppler
signals difficult.  While the transit technique is also biased against
young and therefore highly photometrically variable stars, this effect is
less severe.  To quantify these effects in terms of characteristic mean
age offsets, we calculate the velocity dispersions of three additional
samples of giant exoplanets with $P_{\text{orb}} > 10$ d and plot these
results in Figure \ref{fig:methodscomp}: (1) systems discovered via the
Doppler technique, (2) systems discovered via the transit technique, and
(3) the union of (1) and (2).  As most giant planets with $P_{\text{orb}}
> 10$ d have been discovered via the Doppler technique, sample (3) is
dominated by older, low-jitter stars and has a velocity dispersion similar
to sample (1).  As expected, those samples' velocity dispersions are much
warmer than that of sample (2).  We are therefore justified in focusing
our analysis of hot Jupiter subpopulations on transit discoveries alone.

While we have established in Section \ref{sec:methods} that multiple
mechanisms contribute to the hot Jupiter population, those analyses
alone do not constrain the relative contributions from each mechanism.
To quantify these proportions, we develop a parametric forward model
for the formation and evolution of the hot Jupiter subpopulations
with these proportions as independent variables.  We model the hot
Jupiter population with three formation mechanisms: (1) an early-time,
uniformly-distributing mechanism, (2) an early-time, peak-populating
mechanism, and (3) a late-time, peak-populating mechanism.  To initialize
our host star and hot Jupiter masses, we self-consistently calculate host
star masses for all of the hot Jupiters in the NASA Exoplanet Archive
and use the system's observed inclinations, Doppler semiamplitudes, and
eccentricities to calculate planet masses.  We initialize our orbital
periods in two different ways.  We first initialize the early-time,
uniform component by distributing some fraction $f_{\text{EU}}$ of hot
Jupiters at initial orbital periods distributed as $\mathcal{U}(1,10)$ d.
We next initialize both the early-time and late-time, peak-populating
components with initial orbital periods distributed as $\mathcal{N}(4,
0.5^2)$ d and proportions $f_{\text{EP}}$ and $f_{\text{LP}}$.  We require
$f_{\text{EU}} + f_{\text{EP}} + f_{\text{LP}} = 1$, so only two of these
three variables are independent ($f_{\text{LP}}$ and $f_{\text{EP}}$
in our forward model).

Because newly-formed giant planets may have nonzero eccentricities, for
the early-time, uniformly-distributing mechanism we use a Beta(0.867,
3.03) eccentricity distribution following \citet{Kipping13}.  For the
both the early-time and late-time peak-populating mechanisms we
use the inner planet eccentricity distribution observed in the Mass
Distribution 3 simulation from \citet{Chatterjee08} planet--planet
scattering simulations.  Though core mass has little effect on the
resulting eccentricity distribution predicted from planet--planet
scattering, we select this set of simulations as it permits core masses
larger than 10 M$_\oplus$ that have been inferred for several planets
\citep[e.g.,][]{Buhler16, Welbanks24, Sing24}.  We approximate the
\citet{Chatterjee08} eccentricity distribution as Beta(2.67, 5.67) and
use it to initialize eccentricities for both peak-populating mechanisms.
For both peak-populating mechanisms, we additionally increase the
initial separations of these planets by a factor of $(1-e^2)^{-1}$ such
that they start out still undergoing circularization before arriving
(approximately, assuming conservation of angular momentum) at an orbital
period corresponding to the selected orbital period distributed as
$\mathcal{N}(4, 0.5^2)$ d.  We initialize host stellar obliquities
for both the early-time and late-time peak-populating mechanisms with
a Beta(1.42, 12.8) distribution multiplied by a scaling factor of 180
degrees to approximate the \citet{Chatterjee08} obliquity distribution.

To model the subsequent tidal evolution in our forward model, we use
the best existing constraint on the giant planet specific tidal quality
factor $Q_p' = 3.56 \times 10^5$ derived for Jupiter from observations
of its Galilean satellites \citep{Lainey09}.  We use the tidal evolution
model presented in \citet{Leconte10} in combination with \citet{Amard19}
rotational stellar evolution models to account for tidal evolution.
We interpolate the \citet{Amard19} grid to obtain a rotational stellar
evolution track for each system considered in our forward model.
Because the specific tidal quality factors for main sequence dwarf stars
are poorly constrained, we execute our forward model for four different
values of $Q_\ast'$: $Q_\ast' = 10^{6.0}$, $Q_\ast' = 10^{6.5}$, $Q_\ast'
= 10^{7.0}$, and $Q_\ast' = 10^{7.5}$.

\begin{figure*}
    \centering
    \includegraphics[width=\linewidth]{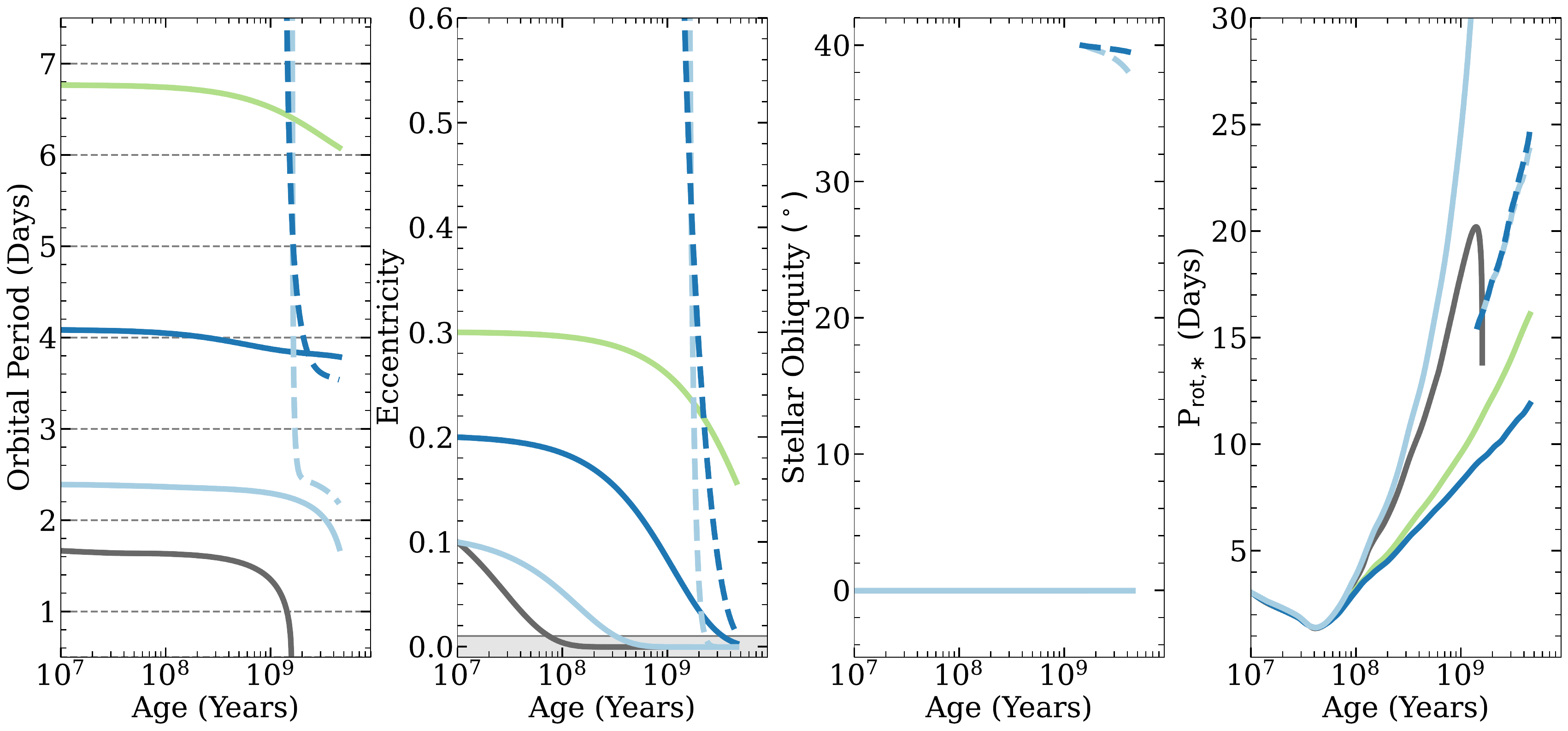}
    \caption{Six possible outcomes of tidal evolution in hot Jupiter
    systems according to the \citet{Leconte10} model assuming the
    rotational stellar evolution models presented in \citet{Amard19}.
    We consider two scenarios for hot Jupiter formation: (1) four
    realizations of an early-arrival mechanism that produces a uniform
    orbital period distribution and (2) two realizations of a late-arrival
    mechanism that produces a peaked orbital period distribution.
    We plot as a function of system age colored lines corresponding to
    system orbital period $P$ (left), eccentricity $e$ (center left),
    stellar obliquity $\psi$ (center right), and stellar rotation period
    $P_{\text{rot}, \ast}$ (right). We indicate the four realizations of
    the early arrival scenario with solid lines and the two realizations
    of the late arrival scenario with dashed lines.  Early-arriving hot
    Jupiter systems with small separations experience inspiral, while all
    hot Jupiter systems experience eccentricity damping to some extent.
    For hot Jupiter systems outside the debiased orbital period peak
    (light green), circularization to $e < 0.01$ as indicated by the gray
    rectangle typically takes longer than the current age of the system.
    Hot Jupiter systems near the debiased orbital period peak (dark blue)
    may circularize on a comparable timescale to the systems' ages,
    while early-arriving small-separation systems circularize quickly
    (light blue) and may be engulfed (gray).  On the other hand, hot
    Jupiter systems formed via a late-time, peak-populating mechanism
    would be less evolved and still observable later in their host stars'
    main sequence lifetimes.  We expect these late arrivals to have
    nonzero host star obliquities, while the host stars of hot Jupiters
    that have experienced significant inspiral should be rotating more
    quickly than stars of similar masses and ages.}
    \label{fig:individual_scenarios}
\end{figure*}

As planetary systems are continuously being formed in the Milky Way's thin
disk, we simulate approximately constant system formation up to the age of
the solar neighborhood by uniformly distributing system formation times as
$\mathcal{U}(0, \text{Min}(t_{\text{MS}}, 8))$ Gyr, where t$_{\text{MS}}$
is the time at which the host star's surface gravity $\log g < 4.0$
(i.e., its main sequence lifetime).  As there is no first principles
reason to favor any particular value for the time between the early-time
and late-time hot Jupiter formation mechanisms, we execute our forward
model using a range of potential time lags.  For the late-time hot Jupiter
formation mechanism, we parameterize this time lag as the average time
between system formation and the formation of the hot Jupiter, $\Delta
\tau$.  We first construct a sequence of values for $\Delta \tau$ from
500 Myr to 2.5 Gyr spaced by 250 Myr.  Any smaller values of $\Delta \tau$
would be difficult to observationally distinguish, and any larger values
of $\Delta \tau$ would be unphysical given our inferred subpopulation
characteristic mean ages.  For the early-time hot Jupiter formation
mechanisms, we start tidal evolution 10 Myr after system formation
corresponding to the dissipation of the parent protoplanetary disk.
For hot Jupiters formed via the late-time, peak-populating mechanism
with $P_{\text{orb}}$ distributed as $\mathcal{N}(4, 0.5^2)$ d, we
start tidal evolution after an additional period of time distributed as
$\mathcal{N}(\Delta \tau, 0.5^2)$ Gyr.  We illustrate several scenarios
for individual hot Jupiter systems under these formation assumptions
and $Q_\ast' = 10^7$ in Figure \ref{fig:individual_scenarios}.

On each iteration of our 500-iteration Monte Carlo simulations, an
initial hot Jupiter population of 500 systems is generated according
to the parameters described above.  It is then evolved to the present,
eliminating systems where the planet is tidally disrupted and ignoring
the few (if any) systems with orbital periods that increase beyond
$P_{\text{orb}} > 10$ d.  We split the surviving population into thirds
by orbital period and calculate each subpopulation's median age.
Marginalizing over all 500 iterations, we calculate the 16th, 50th,
and 84th quantiles of the resulting age distributions.

\begin{figure*}
    \centering
    \includegraphics[width=\linewidth]{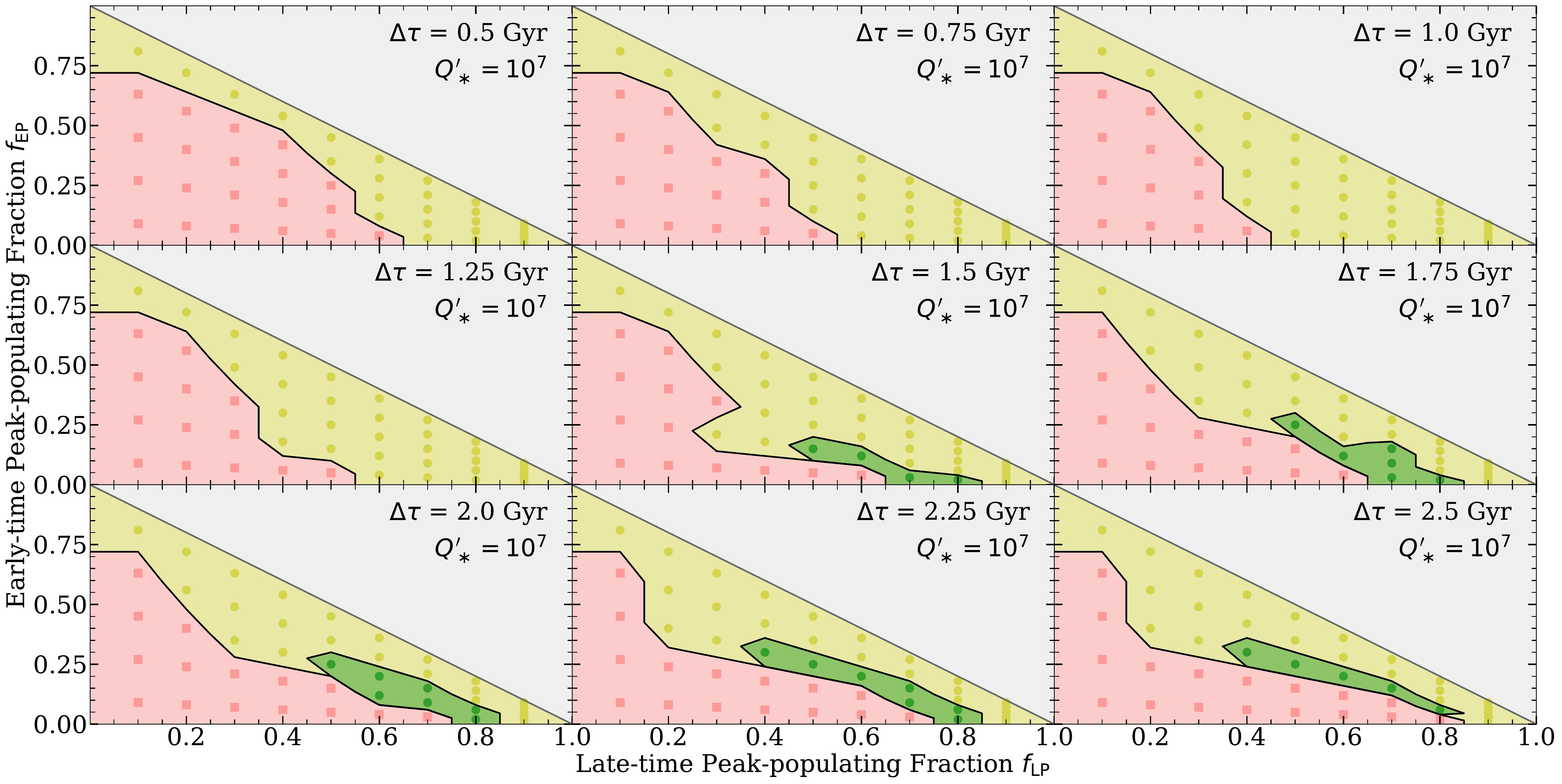}
    \includegraphics[width=\linewidth]{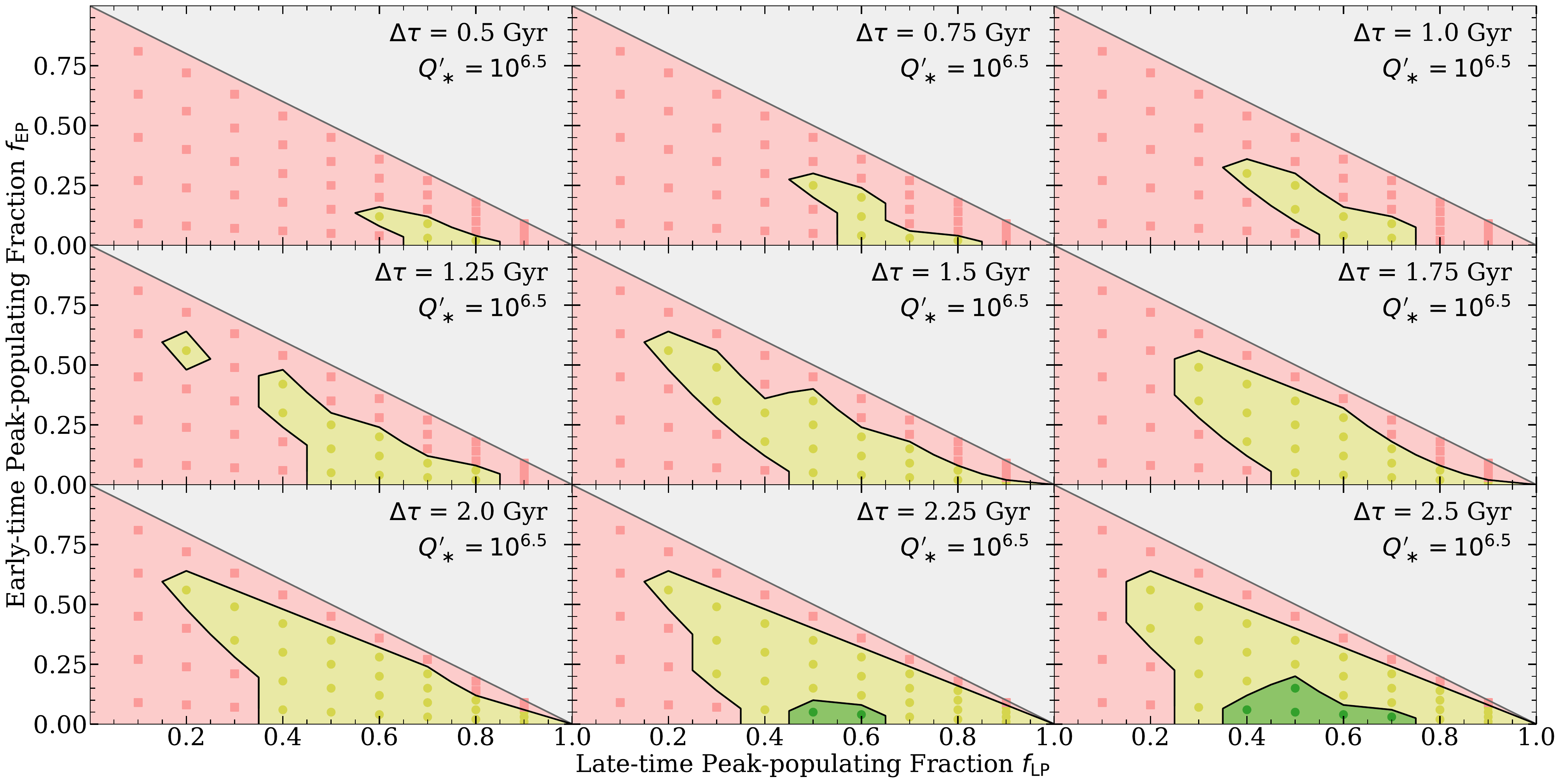}
    \caption{Outcome of our parametric forward model for the formation
    and evolution of the hot Jupiter population assuming stellar tidal
    quality factors $Q'_\ast = 10^7$ (top) and $Q'_\ast = 10^{6.5}$
    (bottom).  In each of the nine panels of each plot we vary four
    parameters in our forward model: (1) the fraction of the hot Jupiter
    population that formed via a late-time, peak-populating mechanism
    $f_{\text{LP}}$, (2) the fraction of the hot Jupiter population
    that formed via an early-time, uniformly-distributing mechanism
    $f_{\text{EU}}$, (3) the fraction of the hot Jupiter population that
    formed via an early-time, peak-populating mechanism $f_{\text{EP}}$,
    and (4) the average time between system formation and late-time
    mechanism occurrence for systems that formed via the late-time
    mechanism $\Delta \tau$.  We require $f_{\text{LP}} + f_{\text{EU}}
    + f_{\text{EP}} = 1$.  To populate each panel, for each point in
    ($f_{\text{LP}}$, $f_{\text{EU}}$, $f_{\text{EP}}$) space we execute
    a Monte Carlo simulation as described in Section \ref{sec:disc}.
    Because we find that the characteristic mean ages of our three hot
    Jupiter subpopulations decrease with orbital period, we plot in red
    regions of parameter space inconsistent with this ordering or with an
    outside-peak subpopulation not significantly younger than the near-
    and inside-peak subpopulations.  In contrast, we plot in yellow and
    green regions of parameter space where both of these criteria are met.
    In green regions it is also true that the characteristic mean age of
    the outside-peak subpopulation is at least 750 Myr younger than than
    the near- and inside-peak subpopulations and that the characteristic
    mean age of the inside-peak subpopulation is at most 200 Myr older
    than the near-peak subpopulation.  We find $0.4 \lesssim f_{\text{LP}}
    \lesssim 0.8$, $\Delta \tau \gtrsim 1.5$ Gyr, and $10^{6.5} \lesssim
    Q_\ast' \lesssim 10^7$ best reproduce our observations, indicating
    that at least 40\% hot Jupiters are late arrivals taking more than
    1.5 Gyr to become hot Jupiters.}
    \label{fig:forwardmodelQ7}
\end{figure*}

\begin{figure*}
    \centering
    \includegraphics[width=\linewidth]{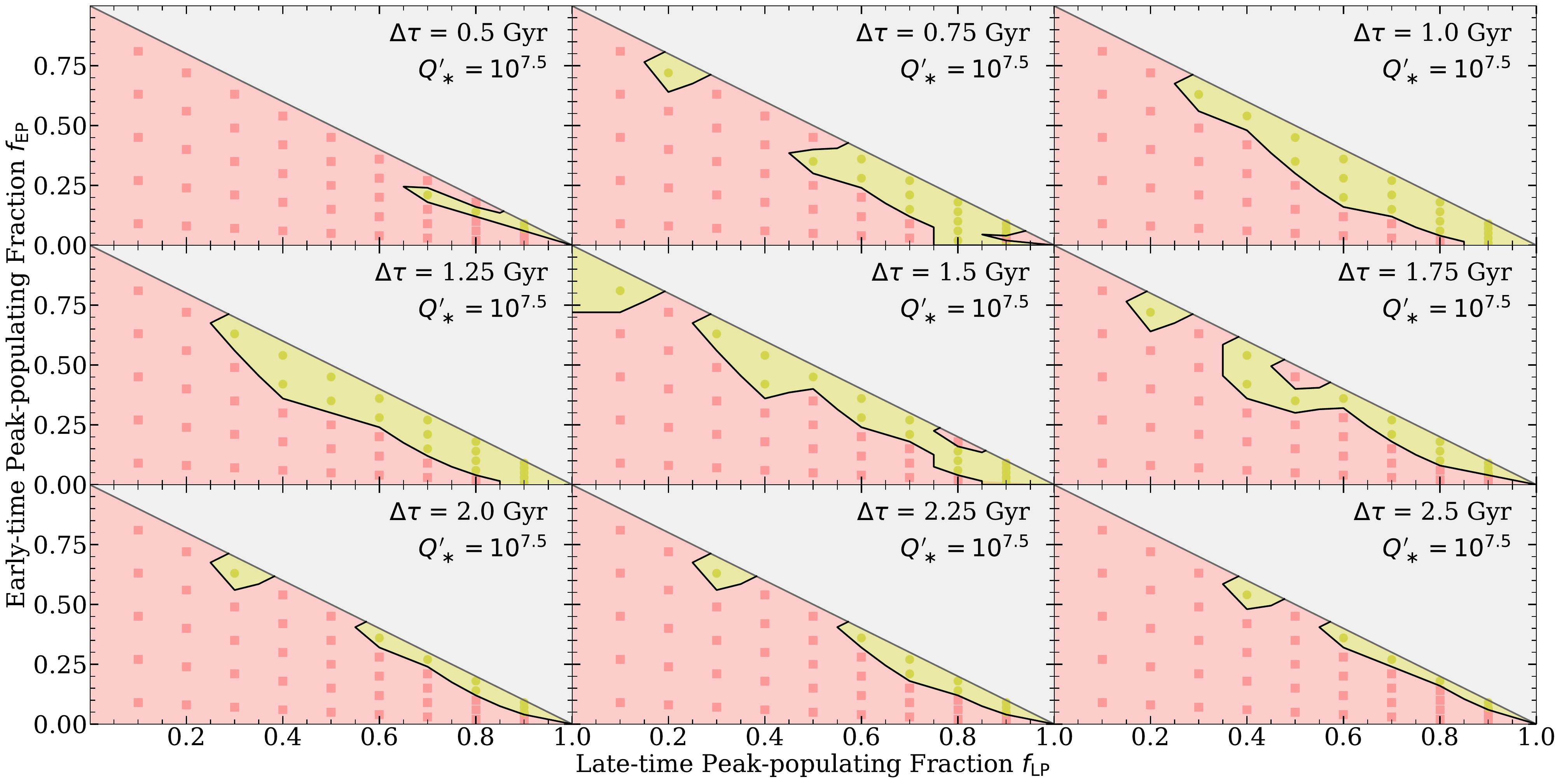}
    \includegraphics[width=\linewidth]{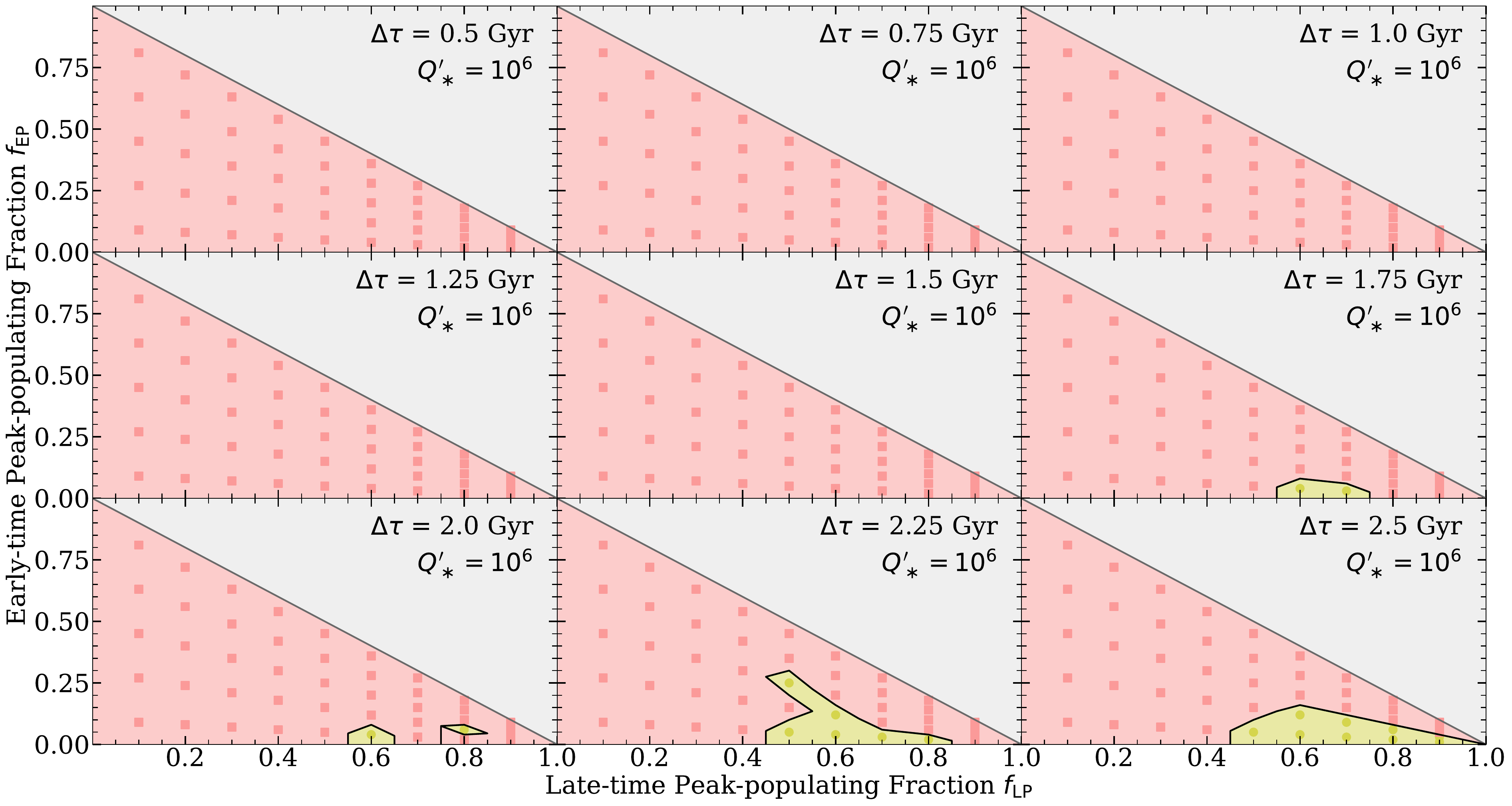}
    \caption{Outcome of our parametric forward model for the formation and
    evolution of the hot Jupiter population assuming stellar tidal quality
    factors $Q'_\ast = 10^{7.5}$ (top) and $Q_\ast' = 10^{6}$ (bottom) and
    therefore more/less dissipative host stars than presented in Figure
    \ref{fig:forwardmodelQ7}.  While $Q'_\ast = 10^{6}$ and $Q'_\ast =
    10^{7.5}$ do not produce outcomes quantitatively consistent with our
    results, they do produce results qualitatively consistent with our
    conclusion that $\gtrsim\!\!40\%$ of the hot Jupiter population are
    late arrivals with $\Delta \tau \approx 2$ Gyr.  We therefore favor
    $Q_\ast'$ values in the range $10^{6.5} \lesssim Q_\ast' \lesssim
    10^7$ because those parameters result in quantitative agreement with
    our observations.  Nevertheless, the range $10^{6} \lesssim Q_\ast'
    \lesssim 10^{7.5}$ is qualitatively consistent with our observations.}
    \label{fig:forwardmodelQ6.5}
\end{figure*}

We plot the final outcome of our forward modeling in Figures
\ref{fig:forwardmodelQ7} and \ref{fig:forwardmodelQ6.5}.  We find that
$10^{6.5}\lesssim Q_\ast' \lesssim 10^7$, 0.4 $\lesssim f_{\text{LP}}
\lesssim$ 0.8, and $\Delta \tau \gtrsim 1.5$ Gyr can quantitatively
reproduce our observations that: (1) subpopulation age decreases with
orbital period, (2) the outside-peak subpopulation is significantly
younger than the near- and inside-peak subpopulations, (3) the
characteristic mean age of the outside-peak subpopulation is at least
750 Myr younger than than the near- and inside-peak subpopulations, and
(4) the characteristic mean age of the inside-peak subpopulation is at
most 200 Myr older than the near-peak subpopulation.  In other words,
we find that at least 40\% of hot Jupiters formed via a late-time,
peak-populating mechanism with a characteristic time delay of more than
1.5 Gyr.  Our conclusion that 0.4 $\lesssim f_{\text{LP}} \lesssim$ 0.8 is
consistent with the \citet{Jackson23} constraint that no more than 62\%
of hot Jupiters in the Kepler field were produced via high-eccentricity
migration followed by tidal circularization.  While we can only exclude
values for $Q_\ast'$ outside the interval $(10^6,10^{7.5})$, we favor
values in the interval $(10^{6.5},10^7)$ because they both produce forward
models quantitatively consistent with our observations and are consistent
with the \citet{Hamer19} constraint $Q_\ast' \lesssim 10^7$.  A broader
range of parameters $Q_\ast'$, $f_{\text{LP}}$, and $\Delta \tau$ can
qualitatively reproduce our observations that: (1) subpopulation age
decreases with orbital period, and (2) the outside-peak subpopulation is
significantly younger than the near- and inside-peak subpopulations. In
this qualitative agreement case, only a minority of hot Jupiters can
form via the early-time, uniformly-populating mechanism that we use to
model disk migration or in situ formation.

While the near- and inside-peak subpopulations of the overall hot
Jupiter population experience significant tidal inspiral during the main
sequence lifetimes of their host stars, the outside-peak subpopulation is
unaffected by inspiral in this time interval.  Therefore, the outside-peak
subpopulation is robust to tidal inspiral for the entire main sequence
lifetimes of their host stars.  It is this outside-peak subpopulation that
was underrepresented in the \citet{Hamer19} analysis that was dominated
by ground-based transit discoveries that has since been observed as hot
Jupiters orbiting evolved stars \citep[e.g.,][]{Saunders22, Saunders25,
Grunblatt22, Pereira24}.  This reconciles the apparently conflicting
claims of \citet{Hamer19} and \citet{Grunblatt23}.

Our observed ages and population modeling imply that most of the hot
Jupiters that exist near the debiased peak of the orbital period
distribution formed in a different manner from those outside
the debiased peak.  The outside-peak hot Jupiter subpopulation
therefore represents the inner edge of the aligned, disk migration/in
situ formation-dominated regime that we expect to produce small
host star obliquities.  This expectation has been confirmed by
Rossiter–McLaughlin  measurements of longer-period giant planet
systems\footnote{See, for example, \citet{Wang21}, \citet{Knudstrup22},
\citet{Rice22b}, \citet{Lubin23}, \citet{Sedaghati23}, \citet{Wright23},
\citet{EspinozaRetamal23}, \citet{Hu24}, \citet{Wang24},
\citet{Bieryla25}, and \citet{EspinozaRetamal25}.}.  Because the
hot Jupiters in the outside-peak subpopulation take much longer
than their current system ages to circularize and therefore exhibit
nonzero eccentricities, our interpretation is also consistent with the
observation that eccentric hot Jupiters' orbits do not exhibit the same
obliquity-host star mass trend \citep{Rice22} as hot Jupiters on circular
orbits \citep{Schlaufman10, Winn10}.

If our interpretation of our observations is correct, then the peak of
the hot Jupiter orbital period distribution should be less prominent
at ages $\tau \lesssim 1$ Gyr.  We therefore predict that as the sample
of hot Jupiters with system ages $\tau \lesssim 1$ Gyr grows, it should
have a less pronounced peak in its orbital period distribution.  We also
predict that the occurrence of wide-separation giant planet, brown dwarf,
or stellar companions should be higher for the near-peak subpopulation
than the outside-peak subpopulation.  We emphasize that these conclusions
have been made possible as a result of TESS's many discoveries of hot
Jupiters in the outside-peak subpopulation that were largely invisible
to ground-based transit surveys.  Therefore, studies of hot Jupiter
demographics that were dominated by ground-based discoveries should be
reevaluated in light of these new discoveries.

In addition to our prediction for the orbital period distribution of
hot Jupiters orbiting stars younger than about 1 Gyr, we also expect
an increased occurrence of additional wide-separation companions.
In contrast to hot Jupiters outside the debiased peak of the orbital
period distribution, hot Jupiters near the debiased orbital period peak
could evince these additional companions (1) in Doppler data as additional
Keplerian orbits or an increased occurrence of long-term trends, (2) in
transit data as an increased occurrence of transit-timing variations, or
(3) in astrometric data as additional Keplerian orbits or an increased
occurrence of astrometric accelerations.  We therefore advocate for
long-term baseline monitoring of relatively recently discovered hot
Jupiters that are part of the outside-peak subpopulation to build up
the time baseline needed to evaluate these possibilities.

These interpretations of our observations could also explain in part
the increased hot Jupiter occurrence in pure Doppler-based surveys
for hot Jupiters \citep[e.g.,][]{Mayor11, Wright12} in comparison
to the diminished hot Jupiter occurrence in transit-based surveys
\citep[e.g.,][]{Santerne12, Fressin13, Santerne16}.  While \citet{Moe21}
persuasively argued that a difference in the binarity of the Doppler-
and transit-surveyed populations plays the primary role in this
apparent occurrence difference, other factors may also contribute.
If as suggested by Figure \ref{fig:methodscomp} Doppler-based surveys
are biased towards older stars, then more Doppler-searched systems
would have had enough time for a hot Jupiter to form as described above
than a similar sample of systems searched for transiting hot Jupiters.
Indeed, Doppler surveyors preferentially target stars with \ion{Ca}{2}
H and K emission indicative of ages greater than 2 Gyr---exactly the
time delay indicated by our forward model \citep[e.g.,][]{Marcy05}.

\begin{figure*}
    \centering
    \includegraphics[width=\linewidth]{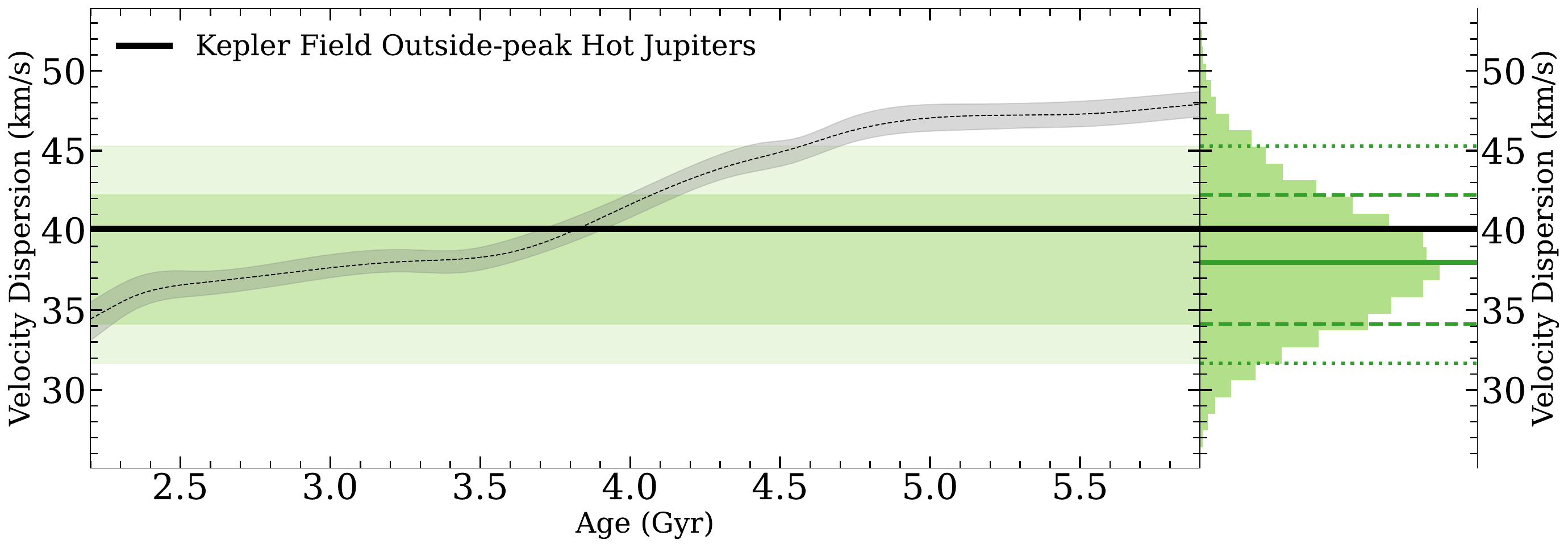}
    \caption{Velocity dispersions of Kepler-discovered outside-peak
    hot Jupiter host stars and matched control samples of Kepler field
    stars without transiting hot Jupiters.  We follow the methodology
    presented in \citet{Hamer19} and calculate velocity dispersions
    for both populations after constructing control samples of Kepler
    field stars matched in mass and metallicity to the host stars of
    Kepler-discovered outside-peak hot Jupiters using the \citet{Berger20}
    catalog of stellar parameters.  We plot as the black dashed line the
    \citet{Schmidt24} Kepler field age--velocity dispersion and indicate
    its 16th/84th interquantile range in gray.  We plot as the heavy black
    line at $\sigma \approx 40$ km s$^{-1}$ the velocity dispersion of
    the sample of Kepler-discovered outside-peak hot Jupiter host stars.
    On the left, we plot the 16th/84th and 5th/95th interquantile ranges
    of the control sample velocity dispersion distribution as the green
    and light green rectangles.  On the right, we plot a histogram
    of individual control sample velocity dispersions and indicate
    its median with a solid green line and its 16th/84th and 5th/95th
    interquantile ranges with horizontal green dashed and dotted lines.
    We find that the velocity dispersion of the sample of host stars
    of Kepler-discovered outside-peak hot Jupiters is fully consistent
    with the velocity dispersion distribution expected from matched
    control samples of Kepler field stars without transiting hot Jupiters.
    This consistency validates our interpretation of the results presented
    in this article as evidence that hot Jupiters outside the debiased
    peak of the hot Jupiter orbital period distribution arrived early in
    the evolution of their planetary systems and are subsequently survive
    tidal interactions while their host stars are on the main sequence.}
    \label{fig:keplertest}
\end{figure*}

If our interpretation of our observations is correct, then the age
of the outside-peak subpopulation should be consistent with the age
of a matched control sample of stars without observed hot Jupiters.
The Kepler field is the ideal environment for this validation of our
interpretation, as the sample of dwarf stars searched for transiting
planets by Kepler is known.  We conduct this test of our proposed scenario
for the formation of the outside-peak subpopulation by comparing (1)
the velocity dispersion of the host stars of the Kepler-discovered hot
Jupiter sample's outside-peak subpopulation to (2) an ensemble of dwarf
stars in the Kepler field searched for planets but without observed
transiting hot Jupiters. Following the approach of \citet{Hamer19}, we
match these samples in mass and metallicity using the \citet{Berger20}
catalog of stellar parameters and confirm that the velocity dispersion
of the outside-peak subpopulation of the Kepler field's hot Jupiter
population is indeed consistent with velocity dispersion distribution
produced by control samples matched in mass and metallicity.  We visualize
the results of this test in Figure \ref{fig:keplertest}.  This analysis
further supports our conclusion that the hot Jupiters in the outside-peak
subpopulation formed early and have not been substantially affected by
tidal inspiral to the point of disruption by their host stars.

\begin{figure*}
    \centering
    \includegraphics[width=\linewidth]{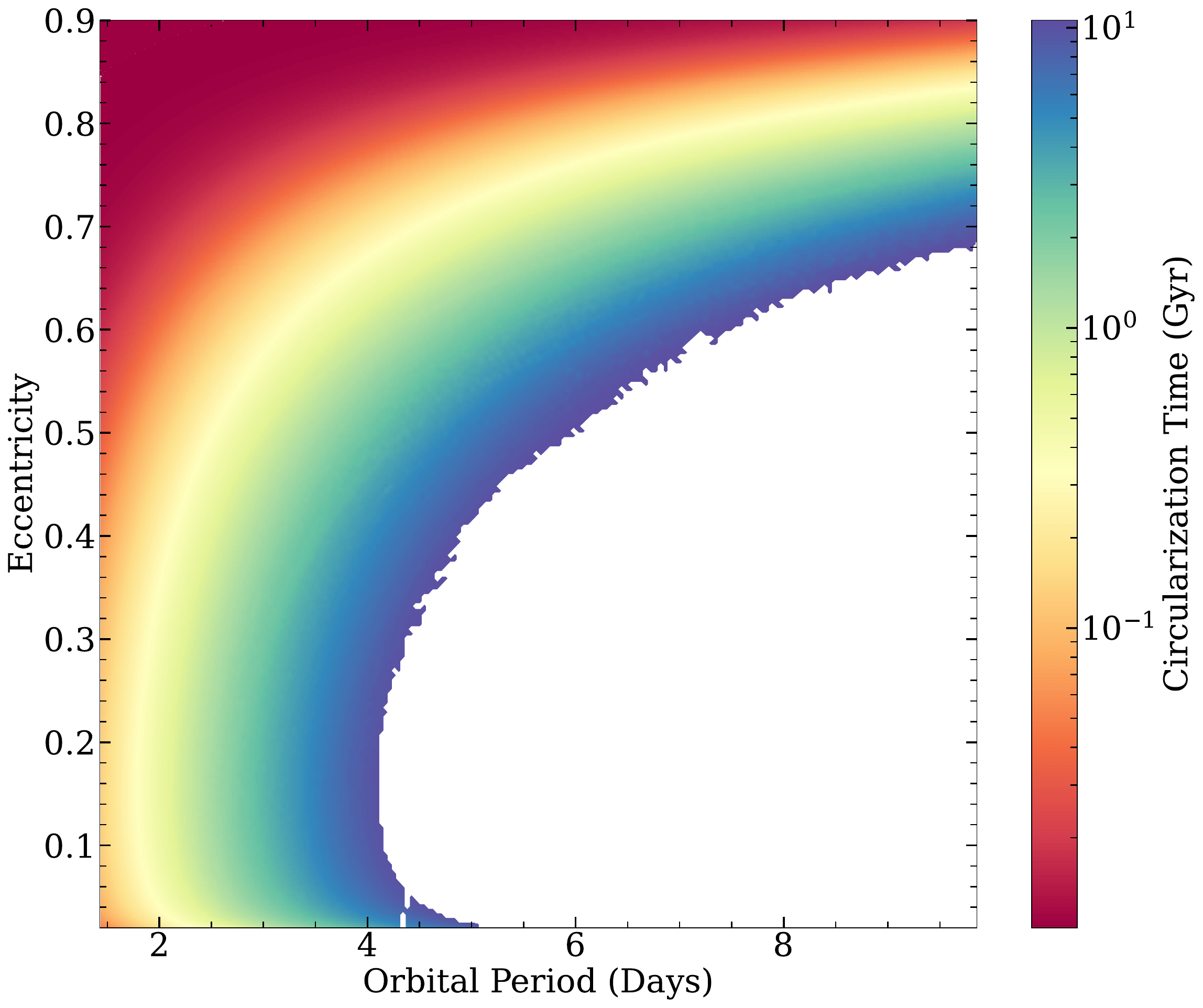}
    \caption{Circularization time as a function of initial orbital
    period and eccentricity for a 1 M$_{\text{Jup}}$ planet orbiting
    a 1 M$_\odot$ star using the \citet{Leconte10} model assuming the
    rotational stellar evolution models presented in \citet{Amard19}
    and a stellar obliquity of zero.  We assume a planet tidal quality
    factor $Q_p' = 3.56 \times 10^5$ \citep{Lainey09} and a stellar
    tidal quality factor $Q_\ast' = 10^7$.  The color bar indicates the
    time for a system's eccentricity to be tidally damped to $e < 0.01$
    at which point it is observationally indistinguishable from $e =
    0$. We indicate in white regions in which $e \geq 0.01$ for more
    than 10 Gyr.  Since the peak of the debiased hot Jupiter orbital
    period distribution is near $P = 4$ d, we expect that many systems
    in our outside-peak subpopulation will not have fully circularized
    by the time their host stars leave the main sequence.}
    \label{fig:circularizationtime}
\end{figure*}

Finally, we note that many objects classified as hot Jupiters can
retain the post-formation orbital eccentricities for many Gyr.
Since obliquity damping generally takes longer than eccentricity
damping, the same is also true for the stellar obliquities of hot
Jupiter host stars.  To demonstrate this, we calculate the eccentricity
damping timescale for a range of hot Jupiter system initial semimajor
axes and eccentricities.  Assuming a median rotation \citet{Amard19}
1 M$_\odot$ stellar model with $Q_\ast' = 10^7$, the \citet{Leconte10}
tidal evolution model suggests that a significant fraction of hot
Jupiters ($P_{\text{orb}} \gtrsim 4$ d and $e \lesssim 0.3$) will avoid
orbit circularization and therefore obliquity damping on Gyr timescales
(Figure \ref{fig:circularizationtime}).  Our calculations show that
studies of the post-formation eccentricities and obliquities of giant
planets need not restrict their samples to $P_{\text{orb}} > 10$ d.

\section{Conclusion} \label{sec:summ}

Several hot Jupiter formation scenarios have been suggested to explain
the existence of the observed peak in the hot Jupiter orbital period
distribution.  We propose that the relative ages of hot Jupiters inside,
near, and outside the peak in the orbital period distribution can be used
to distinguish between these suggested possibilities.  We therefore split
the population of transiting solar neighborhood hot Jupiter systems into
three subpopulations inside, near, and outside the debiased orbital period
peak and then use their Galactic velocity dispersions to infer their
characteristic mean ages.  We find that the outside-peak subpopulation
($4.545 \text{ d} < P_{\text{orb}} < 10$ d) has a characteristic mean age
in the range $2.20 \text{ Gyr} \lesssim \tau \lesssim 2.36$ Gyr.  On the
other hand, we find that the near-peak subpopulation ($3.259 \text{ d} <
P_{\text{orb}} < 4.545$ d) and inside-peak subpopulation ($P_{\text{orb}}
< 3.259$ d) have characteristic mean ages in the ranges $3.11 \text{
Gyr}\lesssim \tau \lesssim 3.27$ Gyr and $3.16 \text{ Gyr} \lesssim \tau
\lesssim 3.30$ Gyr.  These age offsets cannot be explained by differences
in differing stellar mass distributions in each subpopulation.  The older
ages of the near- and inside-peak subpopulations suggest that many
of these systems are formed by a late-time, peak-populating formation
mechanism like high-eccentricity migration.  They are then sculpted by
tidal inspiral over billions of years.

We create a parametric forward model of the evolution of the hot
Jupiter population and find that $Q_\ast'$ values in the range $10^{6.5}
\lesssim Q_\ast' \lesssim 10^7$ quantitatively reproduce our observations.
We further find that between 40\% and 70\% of the hot Jupiter population
must have formed via a late-time, peak-populating mechanism with a
characteristic post-formation timescale that exceeds 1.5 Gyr.  When we
consider more or less dissipative host stars for our model, while
we find that $Q_\ast'$ values in the range $10^{6} \lesssim Q_\ast'
\lesssim 10^{7.5}$ can qualitatively reproduce our observations, they
cannot do so quantitatively.  Our observations imply that most of the
hot Jupiters near the debiased peak in the orbital period distribution
are late arrivals that took more than 1.5 Gyr to become hot Jupiters.
Hot Jupiters inside the debiased orbital period peak formed in a similar
manner but have experienced more dynamically important tidal inspiral.
In contrast, most hot Jupiters outside the debiased orbital peak
formed via disk migration or in situ formation in a similar fashion to
longer-period giant planets.  Those wide-separation systems survive tidal
evolution for the duration of their host stars' main sequence lifetimes
and can be observed transiting evolved post-main sequence stars.

\begin{acknowledgments}

We thank the anonymous referee for their helpful comments.  We thank
Daniel Thorngren for a helpful comment about ODE integration for edge
cases and Nicole Crumpler for a helpful comment about parallelization.
Stephen P.\ Schmidt is supported by the National Science Foundation
Graduate Research Fellowship Program under Grant No. DGE2139757.
This work has made use of data from the European Space Agency
(ESA) mission {\it Gaia} (\url{https://www.cosmos.esa.int/gaia}),
processed by the {\it Gaia} Data Processing and Analysis Consortium
(DPAC, \url{https://www.cosmos.esa.int/web/gaia/dpac/consortium}).
Funding for the DPAC has been provided by national institutions,
in particular the institutions participating in the {\it Gaia}
Multilateral Agreement.  This research has made use of the SIMBAD
database, operated at CDS, Strasbourg, France \citep{Wenger2000}.
This research has made use of the VizieR catalog access tool, CDS,
Strasbourg, France.  The original description of the VizieR service was
published in \citet{VizieR}.  This research has made use of the NASA
Exoplanet Archive \citep{ExoplanetArchive, Christiansen25}, which is
operated by the California Institute of Technology, under contract with
the National Aeronautics and Space Administration under the Exoplanet
Exploration Program.  Funding for SDSS-III has been provided by the
Alfred P. Sloan Foundation, the Participating Institutions, the National
Science Foundation, and the U.S. Department of Energy Office of Science.
The SDSS-III web site is \url{http://www.sdss3.org/}. SDSS-III is
managed by the Astrophysical Research Consortium for the Participating
Institutions of the SDSS-III Collaboration including the University
of Arizona, the Brazilian Participation Group, Brookhaven National
Laboratory, Carnegie Mellon University, University of Florida, the French
Participation Group, the German Participation Group, Harvard University,
the Instituto de Astrofisica de Canarias, the Michigan State/Notre
Dame/JINA Participation Group, Johns Hopkins University, Lawrence
Berkeley National Laboratory, Max Planck Institute for Astrophysics,
Max Planck Institute for Extraterrestrial Physics, New Mexico State
University, New York University, Ohio State University, Pennsylvania
State University, University of Portsmouth, Princeton University, the
Spanish Participation Group, University of Tokyo, University of Utah,
Vanderbilt University, University of Virginia, University of Washington,
and Yale University.  Funding for the Sloan Digital Sky Survey IV has
been provided by the Alfred P. Sloan Foundation, the U.S.  Department
of Energy Office of Science, and the Participating Institutions.
SDSS-IV acknowledges support and resources from the Center for High
Performance Computing at the University of Utah.  The SDSS website
is www.sdss4.org.  SDSS-IV is managed by the Astrophysical Research
Consortium for the Participating Institutions of the SDSS Collaboration
including the Brazilian Participation Group, the Carnegie Institution
for Science, Carnegie Mellon University, Center for Astrophysics |
Harvard \& Smithsonian, the Chilean Participation Group, the French
Participation Group, Instituto de Astrof\'isica de Canarias, The Johns
Hopkins University, Kavli Institute for the Physics and Mathematics of
the Universe (IPMU) / University of Tokyo, the Korean Participation
Group, Lawrence Berkeley National Laboratory, Leibniz Institut f\"ur
Astrophysik Potsdam (AIP),  Max-Planck-Institut f\"ur Astronomie (MPIA
Heidelberg), Max-Planck-Institut f\"ur Astrophysik (MPA Garching),
Max-Planck-Institut f\"ur Extraterrestrische Physik (MPE), National
Astronomical Observatories of China, New Mexico State University, New
York University, University of Notre Dame, Observat\'ario Nacional /
MCTI, The Ohio State University, Pennsylvania State University, Shanghai
Astronomical Observatory, United Kingdom Participation Group, Universidad
Nacional Aut\'onoma de M\'exico, University of Arizona, University
of Colorado Boulder, University of Oxford, University of Portsmouth,
University of Utah, University of Virginia, University of Washington,
University of Wisconsin, Vanderbilt University, and Yale University.
This research has made use of NASA's Astrophysics Data System.

\end{acknowledgments}

\vspace{5mm}
\facilities{ADS, CDS, Du Pont (APOGEE), Exoplanet Archive, Gaia, Sloan
(APOGEE)}

\software{\texttt{astropy} \citep{AstropyI, AstropyII, AstropyIII},
\texttt{astroquery} \citep{astroquery},
\texttt{gaiadr3\_zeropoint} \citep{lin21a},
\texttt{gala} \citep{gala, adrian_price_whelan_2020_4159870},
\texttt{matplotlib} \citep{hunter2007matplotlib},
\texttt{numpy} \citep{harris2020array},
\texttt{pandas} \citep{McKinney_2010, reback2020pandas},
\texttt{pyia} \citep{adrian_price_whelan_2018_1228136},
\texttt{scipy} \citep{jones2001scipy,2020SciPy-NMeth},
\texttt{TOPCAT} \citep{TOPCAT}}

\bibliography{article_bibliography}{}

@ARTICLE{Nataf24,
       author = {{Nataf}, David M. and {Schlaufman}, Kevin C. and {Reggiani}, Henrique and {Hahn}, Isabel},
        title = "{Accurate, Precise, and Physically Self-consistent Ages and Metallicities for 400,000 Solar Neighborhood Subgiant Branch Stars}",
      journal = {\apj},
         year = 2024,
        month = nov,
       volume = {976},
       number = {1},
          eid = {87},
        pages = {87},
          doi = {10.3847/1538-4357/ad7c4e},
archivePrefix = {arXiv},
       eprint = {2407.18307},
 primaryClass = {astro-ph.SR},
       adsurl = {https://ui.adsabs.harvard.edu/abs/2024ApJ...976...87N}
}

@ARTICLE{Schlaufman18,
       author = {{Schlaufman}, Kevin C.},
        title = "{Evidence of an Upper Bound on the Masses of Planets and Its Implications for Giant Planet Formation}",
      journal = {\apj},
         year = 2018,
        month = jan,
       volume = {853},
       number = {1},
          eid = {37},
        pages = {37},
          doi = {10.3847/1538-4357/aa961c},
archivePrefix = {arXiv},
       eprint = {1801.06185},
 primaryClass = {astro-ph.EP},
       adsurl = {https://ui.adsabs.harvard.edu/abs/2018ApJ...853...37S}
}

@ARTICLE{Schmidt24,
       author = {{Schmidt}, Stephen P. and {Schlaufman}, Kevin C. and {Hamer}, Jacob H.},
        title = "{Resonant and Ultra-short-period Planet Systems Are at Opposite Ends of the Exoplanet Age Distribution}",
      journal = {\aj},
         year = 2024,
        month = sep,
       volume = {168},
       number = {3},
          eid = {109},
        pages = {109},
          doi = {10.3847/1538-3881/ad5d76},
archivePrefix = {arXiv},
       eprint = {2407.04765},
 primaryClass = {astro-ph.EP},
       adsurl = {https://ui.adsabs.harvard.edu/abs/2024AJ....168..109S}
}

@ARTICLE{Hamer19,
       author = {{Hamer}, Jacob H. and {Schlaufman}, Kevin C.},
        title = "{Hot Jupiters Are Destroyed by Tides While Their Host Stars Are on the Main Sequence}",
      journal = {\aj},
         year = 2019,
        month = nov,
       volume = {158},
       number = {5},
          eid = {190},
        pages = {190},
          doi = {10.3847/1538-3881/ab3c56},
archivePrefix = {arXiv},
       eprint = {1908.06998},
 primaryClass = {astro-ph.EP},
       adsurl = {https://ui.adsabs.harvard.edu/abs/2019AJ....158..190H}
}

@ARTICLE{Hamer20,
       author = {{Hamer}, Jacob H. and {Schlaufman}, Kevin C.},
        title = "{Ultra-short-period Planets Are Stable against Tidal Inspiral}",
      journal = {\aj},
         year = 2020,
        month = sep,
       volume = {160},
       number = {3},
          eid = {138},
        pages = {138},
          doi = {10.3847/1538-3881/aba74f},
archivePrefix = {arXiv},
       eprint = {2007.10944},
 primaryClass = {astro-ph.EP},
       adsurl = {https://ui.adsabs.harvard.edu/abs/2020AJ....160..138H}
}

@ARTICLE{Holman97,
       author = {{Holman}, Matthew and {Touma}, Jihad and {Tremaine}, Scott},
        title = "{Chaotic variations in the eccentricity of the planet orbiting 16 Cygni B}",
      journal = {\nat},
         year = 1997,
        month = mar,
       volume = {386},
       number = {6622},
        pages = {254-256},
          doi = {10.1038/386254a0},
       adsurl = {https://ui.adsabs.harvard.edu/abs/1997Natur.386..254H}
}

@ARTICLE{Mazeh97,
       author = {{Mazeh}, Tsevi and {Krymolowski}, Yuval and {Rosenfeld}, Gady},
        title = "{The High Eccentricity of the Planet Orbiting 16 Cygni B}",
      journal = {\apjl},
         year = 1997,
        month = mar,
       volume = {477},
       number = {2},
        pages = {L103-L106},
          doi = {10.1086/310536},
archivePrefix = {arXiv},
       eprint = {astro-ph/9611135},
 primaryClass = {astro-ph},
       adsurl = {https://ui.adsabs.harvard.edu/abs/1997ApJ...477L.103M}
}

@ARTICLE{Christiansen25,
       author = {{Christiansen}, Jessie L. and {McElroy}, Douglas L. and {Harbut}, Marcy and {Ciardi}, David R. and {Crane}, Megan and {Good}, John and {Hardegree-Ullman}, Kevin K. and {Kesseli}, Aurora Y. and {Lund}, Michael B. and {Lynn}, Meca and {Muthiar}, Ananda and {Nilsson}, Ricky and {Oluyide}, Toba and {Papin}, Michael and {Rivera}, Amalia and {Swain}, Melanie and {Susemiehl}, Nicholas D. and {Tam}, Raymond and {van Eyken}, Julian and {Beichman}, Charles},
        title = "{The NASA Exoplanet Archive and Exoplanet Follow-up Observing Program: Data, Tools, and Usage}",
      journal = {\psj},
         year = 2025,
        month = aug,
       volume = {6},
       number = {8},
          eid = {186},
        pages = {186},
          doi = {10.3847/PSJ/ade3c2},
archivePrefix = {arXiv},
       eprint = {2506.03299},
 primaryClass = {astro-ph.EP},
       adsurl = {https://ui.adsabs.harvard.edu/abs/2025PSJ.....6..186C}
}

@ARTICLE{Marcy05,
       author = {{Marcy}, Geoffrey W. and {Butler}, R. Paul and {Vogt}, Steven S. and {Fischer}, Debra A. and {Henry}, Gregory W. and {Laughlin}, Greg and {Wright}, Jason T. and {Johnson}, John A.},
        title = "{Five New Extrasolar Planets}",
      journal = {\apj},
         year = 2005,
        month = jan,
       volume = {619},
       number = {1},
        pages = {570-584},
          doi = {10.1086/426384},
       adsurl = {https://ui.adsabs.harvard.edu/abs/2005ApJ...619..570M}
}

@ARTICLE{Santos04,
       author = {{Santos}, N.~C. and {Israelian}, G. and {Mayor}, M.},
        title = "{Spectroscopic [Fe/H] for 98 extra-solar planet-host stars.  Exploring the probability of planet formation}",
      journal = {\aap},
         year = 2004,
        month = mar,
       volume = {415},
        pages = {1153-1166},
          doi = {10.1051/0004-6361:20034469},
archivePrefix = {arXiv},
       eprint = {astro-ph/0311541},
 primaryClass = {astro-ph},
       adsurl = {https://ui.adsabs.harvard.edu/abs/2004A&A...415.1153S}
}

@ARTICLE{Moe21,
       author = {{Moe}, Maxwell and {Kratter}, Kaitlin M.},
        title = "{Impact of binary stars on planet statistics - I. Planet occurrence rates and trends with stellar mass}",
      journal = {\mnras},
         year = 2021,
        month = nov,
       volume = {507},
       number = {3},
        pages = {3593-3611},
          doi = {10.1093/mnras/stab2328},
archivePrefix = {arXiv},
       eprint = {1912.01699},
 primaryClass = {astro-ph.EP},
       adsurl = {https://ui.adsabs.harvard.edu/abs/2021MNRAS.507.3593M}
}

@ARTICLE{Yee25,
       author = {{Yee}, Samuel W. and {Winn}, Joshua N. and {Hartman}, Joel D. and {Rodriguez}, Joseph E. and {Zhou}, George and {Latham}, David W. and {Quinn}, Samuel N. and {Bieryla}, Allyson and {Collins}, Karen A. and {Eastman}, Jason D. and {Collins}, Kevin I. and {Conti}, Dennis M. and {Jensen}, Eric L.~N. and {Baker}, David and {Barkaoui}, Khalid and {Ba{\textcommabelow s}t{\"u}rk}, {\"O}zg{\"u}r and {Battley}, Matthew P. and {Bayliss}, Daniel and {Beatty}, Thomas G. and {Beletsky}, Yuri and {Belinski}, Alexander A. and {Benkhaldoun}, Zouhair and {Benni}, Paul and {Bosch-Cabot}, Pau and {Brice{\~n}o}, C{\'e}sar and {Brudny}, Andrzej and {Burleigh}, Matthew R. and {Butler}, R. Paul and {Chairetas}, Stavros and {Chontos}, Ashley and {Christiansen}, Jessie and {Ciardi}, David R. and {Clark}, Catherine A. and {Cloutier}, Ryan and {Craig}, Matthew W. and {Crane}, Jeffrey D. and {Dowling}, Nicholas and {Dressing}, Courtney D. and {Emmanuel}, Jehin and {Evans}, Phil and {Everett}, Mark E. and {Fern{\'a}ndez-Rodr{\'\i}guez}, Gareb and {Fern{\'a}ndez Fern{\'a}ndez}, Jorge and {For{\'e}s-Toribio}, Raquel and {Fortenbach}, Charles D. and {Fukui}, Akihiko and {Furlan}, Elise and {Gan}, Tianjun and {Ghachoui}, Mourad and {Giacalone}, Steven and {Gill}, Samuel and {Gillon}, Micha{\"e}l and {Hall}, Kylie and {Hayashi}, Yuya and {Hedges}, Christina and {Higuera}, Jesus and {Hintz}, Eric G. and {Hirsch}, Lea and {Holcomb}, Rae and {Horne}, Keith and {Grau Horta}, Ferran and {Howard}, Andrew W. and {Howell}, Steve B. and {Isaacson}, Howard and {Jenkins}, Jon M. and {Kagetani}, Taiki and {Kamler}, Jacob and {Kendall}, Alicia and {Korth}, Judith and {Kroft}, Maxwell A. and {Lacedelli}, Gaia and {Laloum}, Didier and {Law}, Nicholas and {de Leon}, Jerome Pitogo and {Levine}, Alan M. and {Lewin}, Pablo and {Logsdon}, Sarah E. and {Lund}, Michael B. and {Madsen}, Madelyn M. and {Mann}, Andrew W. and {Mann}, Christopher R. and {Maslennikova}, Nataliia A. and {Matutano}, Sandra and {McCormack}, Mason and {McLeod}, Kim K. and {Michaels}, Edward J. and {Mireles}, Ismael and {Mori}, Mayuko and {Mu{\~n}oz}, Jose A. and {Murgas}, Felipe and {Narita}, Norio and {O'Brien}, Sean M. and {Odden}, Caroline and {Palle}, Enric and {Patel}, Yatrik G. and {Plavchan}, Peter and {Polanski}, Alex S. and {Popowicz}, Adam and {Radford}, Don J. and {Reed}, Phillip A. and {Relles}, Howard M. and {Rice}, Malena and {Ricker}, George R. and {Safonov}, Boris S. and {Savel}, Arjun B. and {Schulte}, Jack and {Schwarz}, Richard P. and {Schweiker}, Heidi and {Seager}, Sara and {Sefako}, Ramotholo and {Shectman}, Stephen A. and {Shporer}, Avi and {Stephens}, Denise C. and {Stockdale}, Chris and {Striegel}, Stephanie and {Tan}, Thiam-Guan and {Teske}, Johanna K. and {Timmermans}, Mathilde and {Ulmer-Moll}, Sol{\`e}ne and {Wang}, Gavin and {Wheatley}, Peter J. and {Yalcinkaya}, Sel{\c{c}}uk and {Zambelli}, Roberto and {Van Zandt}, Judah and {Ziegler}, Carl},
        title = "{The TESS Grand Unified Hot Jupiter Survey. III. Thirty More Giant Planets}",
      journal = {\apjs},
         year = 2025,
        month = sep,
       volume = {280},
       number = {1},
          eid = {30},
        pages = {30},
          doi = {10.3847/1538-4365/aded0d},
archivePrefix = {arXiv},
       eprint = {2507.01855},
 primaryClass = {astro-ph.EP},
       adsurl = {https://ui.adsabs.harvard.edu/abs/2025ApJS..280...30Y}
}

@ARTICLE{Kiseleva98,
       author = {{Kiseleva}, L.~G. and {Eggleton}, P.~P. and {Mikkola}, S.},
        title = "{Tidal friction in triple stars}",
      journal = {\mnras},
         year = 1998,
        month = oct,
       volume = {300},
       number = {1},
        pages = {292-302},
          doi = {10.1046/j.1365-8711.1998.01903.x},
       adsurl = {https://ui.adsabs.harvard.edu/abs/1998MNRAS.300..292K}
}

@ARTICLE{Hamer22,
       author = {{Hamer}, Jacob H. and {Schlaufman}, Kevin C.},
        title = "{Evidence for the Late Arrival of Hot Jupiters in Systems with High Host-star Obliquities}",
      journal = {\aj},
         year = 2022,
        month = jul,
       volume = {164},
       number = {1},
          eid = {26},
        pages = {26},
          doi = {10.3847/1538-3881/ac69ef},
archivePrefix = {arXiv},
       eprint = {2205.00040},
 primaryClass = {astro-ph.EP},
       adsurl = {https://ui.adsabs.harvard.edu/abs/2022AJ....164...26H}
}

@ARTICLE{Hamer24,
       author = {{Hamer}, Jacob H. and {Schlaufman}, Kevin C.},
        title = "{Kepler-discovered Multiple-planet Systems near Period Ratios Suggestive of Mean-motion Resonances Are Young}",
      journal = {\aj},
         year = 2024,
        month = feb,
       volume = {167},
       number = {2},
          eid = {55},
        pages = {55},
          doi = {10.3847/1538-3881/ad110e},
archivePrefix = {arXiv},
       eprint = {2312.02260},
 primaryClass = {astro-ph.EP},
       adsurl = {https://ui.adsabs.harvard.edu/abs/2024AJ....167...55H}
}

@ARTICLE{Fischer05,
       author = {{Fischer}, Debra A. and {Valenti}, Jeff},
        title = "{The Planet-Metallicity Correlation}",
      journal = {\apj},
         year = 2005,
        month = apr,
       volume = {622},
       number = {2},
        pages = {1102-1117},
          doi = {10.1086/428383},
       adsurl = {https://ui.adsabs.harvard.edu/abs/2005ApJ...622.1102F}
}

@ARTICLE{Yee22,
       author = {{Yee}, Samuel W. and {Winn}, Joshua N. and {Hartman}, Joel D. and {Rodriguez}, Joseph E. and {Zhou}, George and {Quinn}, Samuel N. and {Latham}, David W. and {Bieryla}, Allyson and {Collins}, Karen A. and {Addison}, Brett C. and {Angelo}, Isabel and {Barkaoui}, Khalid and {Benni}, Paul and {Boyle}, Andrew W. and {Brahm}, Rafael and {Butler}, R. Paul and {Ciardi}, David R. and {Collins}, Kevin I. and {Conti}, Dennis M. and {Crane}, Jeffrey D. and {Dai}, Fei and {Dressing}, Courtney D. and {Eastman}, Jason D. and {Essack}, Zahra and {For{\'e}s-Toribio}, Raquel and {Furlan}, Elise and {Gan}, Tianjun and {Giacalone}, Steven and {Gill}, Holden and {Girardin}, Eric and {Henning}, Thomas and {Henze}, Christopher E. and {Hobson}, Melissa J. and {Horner}, Jonathan and {Howard}, Andrew W. and {Howell}, Steve B. and {Huang}, Chelsea X. and {Isaacson}, Howard and {Jenkins}, Jon M. and {Jensen}, Eric L.~N. and {Jord{\'a}n}, Andr{\'e}s and {Kane}, Stephen R. and {Kielkopf}, John F. and {Lasota}, Slawomir and {Levine}, Alan M. and {Lubin}, Jack and {Mann}, Andrew W. and {Massey}, Bob and {McLeod}, Kim K. and {Mengel}, Matthew W. and {Mu{\~n}oz}, Jose A. and {Murgas}, Felipe and {Palle}, Enric and {Plavchan}, Peter and {Popowicz}, Adam and {Radford}, Don J. and {Ricker}, George R. and {Rowden}, Pamela and {Safonov}, Boris S. and {Savel}, Arjun B. and {Schwarz}, Richard P. and {Seager}, S. and {Sefako}, Ramotholo and {Shporer}, Avi and {Srdoc}, Gregor and {Strakhov}, Ivan S. and {Teske}, Johanna K. and {Tinney}, C.~G. and {Tyler}, Dakotah and {Wittenmyer}, Robert A. and {Zhang}, Hui and {Ziegler}, Carl},
        title = "{The TESS Grand Unified Hot Jupiter Survey. I. Ten TESS Planets}",
      journal = {\aj},
         year = 2022,
        month = aug,
       volume = {164},
       number = {2},
          eid = {70},
        pages = {70},
          doi = {10.3847/1538-3881/ac73ff},
archivePrefix = {arXiv},
       eprint = {2205.09728},
 primaryClass = {astro-ph.EP},
       adsurl = {https://ui.adsabs.harvard.edu/abs/2022AJ....164...70Y}
}

@ARTICLE{Yee23,
       author = {{Yee}, Samuel W. and {Winn}, Joshua N. and {Hartman}, Joel D. and {Bouma}, Luke G. and {Zhou}, George and {Quinn}, Samuel N. and {Latham}, David W. and {Bieryla}, Allyson and {Rodriguez}, Joseph E. and {Collins}, Karen A. and {Alfaro}, Owen and {Barkaoui}, Khalid and {Beard}, Corey and {Belinski}, Alexander A. and {Benkhaldoun}, Zouhair and {Benni}, Paul and {Bernacki}, Krzysztof and {Boyle}, Andrew W. and {Butler}, R. Paul and {Caldwell}, Douglas A. and {Chontos}, Ashley and {Christiansen}, Jessie L. and {Ciardi}, David R. and {Collins}, Kevin I. and {Conti}, Dennis M. and {Crane}, Jeffrey D. and {Daylan}, Tansu and {Dressing}, Courtney D. and {Eastman}, Jason D. and {Essack}, Zahra and {Evans}, Phil and {Everett}, Mark E. and {Fajardo-Acosta}, Sergio and {For{\'e}s-Toribio}, Raquel and {Furlan}, Elise and {Ghachoui}, Mourad and {Gillon}, Micha{\"e}l and {Hellier}, Coel and {Helm}, Ian and {Howard}, Andrew W. and {Howell}, Steve B. and {Isaacson}, Howard and {Jehin}, Emmanuel and {Jenkins}, Jon M. and {Jensen}, Eric L.~N. and {Kielkopf}, John F. and {Laloum}, Didier and {Leonhardes-Barboza}, Naunet and {Lewin}, Pablo and {Logsdon}, Sarah E. and {Lubin}, Jack and {Lund}, Michael B. and {MacDougall}, Mason G. and {Mann}, Andrew W. and {Maslennikova}, Natalia A. and {Massey}, Bob and {McLeod}, Kim K. and {Mu{\~n}oz}, Jose A. and {Newman}, Patrick and {Orlov}, Valeri and {Plavchan}, Peter and {Popowicz}, Adam and {Pozuelos}, Francisco J. and {Pritchard}, Tyler A. and {Radford}, Don J. and {Reefe}, Michael and {Ricker}, George R. and {Rudat}, Alexander and {Safonov}, Boris S. and {Schwarz}, Richard P. and {Schweiker}, Heidi and {Scott}, Nicholas J. and {Seager}, S. and {Shectman}, Stephen A. and {Stockdale}, Chris and {Tan}, Thiam-Guan and {Teske}, Johanna K. and {Thomas}, Neil B. and {Timmermans}, Mathilde and {Vanderspek}, Roland and {Vermilion}, David and {Watanabe}, David and {Weiss}, Lauren M. and {West}, Richard G. and {Van Zandt}, Judah and {Zejmo}, Michal and {Ziegler}, Carl},
        title = "{The TESS Grand Unified Hot Jupiter Survey. II. Twenty New Giant Planets}",
      journal = {\apjs},
         year = 2023,
        month = mar,
       volume = {265},
       number = {1},
          eid = {1},
        pages = {1},
          doi = {10.3847/1538-4365/aca286},
archivePrefix = {arXiv},
       eprint = {2210.15473},
 primaryClass = {astro-ph.EP},
       adsurl = {https://ui.adsabs.harvard.edu/abs/2023ApJS..265....1Y}
}

@ARTICLE{Schulte24,
       author = {{Schulte}, Jack and {Rodriguez}, Joseph E. and {Bieryla}, Allyson and {Quinn}, Samuel N. and {Collins}, Karen A. and {Yee}, Samuel W. and {Nine}, Andrew C. and {Soares-Furtado}, Melinda and {Latham}, David W. and {Eastman}, Jason D. and {Barkaoui}, Khalid and {Ciardi}, David R. and {Dragomir}, Diana and {Everett}, Mark E. and {Giacalone}, Steven and {Mireles}, Ismael and {Murgas}, Felipe and {Narita}, Norio and {Shporer}, Avi and {Strakhov}, Ivan A. and {Striegel}, Stephanie and {Va{\v{n}}ko}, Martin and {Vowell}, Noah and {Wang}, Gavin and {Ziegler}, Carl and {Bellaver}, Michael and {Benni}, Paul and {Bergeron}, Serge and {Boffin}, Henri M.~J. and {Brice{\~n}o}, C{\'e}sar and {Clark}, Catherine A. and {Collins}, Kevin I. and {de Leon}, Jerome P. and {Dressing}, Courtney D. and {Evans}, Phil and {Esparza-Borges}, Emma and {Fedewa}, Jeremy and {Fukui}, Akihiko and {Gan}, Tianjun and {Gerasimov}, Ivan S. and {Hartman}, Joel D. and {Gill}, Holden and {Gillon}, Micha{\"e}l and {Horne}, Keith and {Horta}, Ferran Grau and {Howell}, Steve B. and {Isogai}, Keisuke and {Jehin}, Emmanu{\"e}l and {Jenkins}, Jon M. and {Karjalainen}, Raine and {Kielkopf}, John F. and {Lester}, Kathryn V. and {Littlefield}, Colin and {Lund}, Michael B. and {Mann}, Andrew W. and {McCormack}, Mason and {Michaels}, Edward J. and {Painter}, Shane and {Palle}, Enric and {Parviainen}, Hannu and {Peterson}, David-Michael and {Pozuelos}, Francisco J. and {Raup}, Zachary and {Reed}, Phillip and {Relles}, Howard M. and {Ricker}, George R. and {Savel}, Arjun B. and {Schwarz}, Richard P. and {Seager}, S. and {Sefako}, Ramotholo and {Srdoc}, Gregor and {Stockdale}, Chris and {Sullivan}, Hannah and {Timmermans}, Mathilde and {Winn}, Joshua N.},
        title = "{Migration and Evolution of giant ExoPlanets (MEEP). I. Nine Newly Confirmed Hot Jupiters from the TESS Mission}",
      journal = {\aj},
         year = 2024,
        month = jul,
       volume = {168},
       number = {1},
          eid = {32},
        pages = {32},
          doi = {10.3847/1538-3881/ad4a57},
archivePrefix = {arXiv},
       eprint = {2401.05923},
 primaryClass = {astro-ph.EP},
       adsurl = {https://ui.adsabs.harvard.edu/abs/2024AJ....168...32S}
}

@INPROCEEDINGS{Ricker14,
       author = {{Ricker}, George R. and {Winn}, Joshua N. and {Vanderspek}, Roland and {Latham}, David W. and {Bakos}, G{\'a}sp{\'a}r. {\'A}. and {Bean}, Jacob L. and {Berta-Thompson}, Zachory K. and {Brown}, Timothy M. and {Buchhave}, Lars and {Butler}, Nathaniel R. and {Butler}, R. Paul and {Chaplin}, William J. and {Charbonneau}, David and {Christensen-Dalsgaard}, J{\o}rgen and {Clampin}, Mark and {Deming}, Drake and {Doty}, John and {De Lee}, Nathan and {Dressing}, Courtney and {Dunham}, E.~W. and {Endl}, Michael and {Fressin}, Francois and {Ge}, Jian and {Henning}, Thomas and {Holman}, Matthew J. and {Howard}, Andrew W. and {Ida}, Shigeru and {Jenkins}, Jon and {Jernigan}, Garrett and {Johnson}, John A. and {Kaltenegger}, Lisa and {Kawai}, Nobuyuki and {Kjeldsen}, Hans and {Laughlin}, Gregory and {Levine}, Alan M. and {Lin}, Douglas and {Lissauer}, Jack J. and {MacQueen}, Phillip and {Marcy}, Geoffrey and {McCullough}, P.~R. and {Morton}, Timothy D. and {Narita}, Norio and {Paegert}, Martin and {Palle}, Enric and {Pepe}, Francesco and {Pepper}, Joshua and {Quirrenbach}, Andreas and {Rinehart}, S.~A. and {Sasselov}, Dimitar and {Sato}, Bun'ei and {Seager}, Sara and {Sozzetti}, Alessandro and {Stassun}, Keivan G. and {Sullivan}, Peter and {Szentgyorgyi}, Andrew and {Torres}, Guillermo and {Udry}, Stephane and {Villasenor}, Joel},
        title = "{Transiting Exoplanet Survey Satellite (TESS)}",
    booktitle = {Space Telescopes and Instrumentation 2014: Optical, Infrared, and Millimeter Wave},
         year = 2014,
       editor = {{Oschmann}, Jr., Jacobus M. and {Clampin}, Mark and {Fazio}, Giovanni G. and {MacEwen}, Howard A.},
       series = {Society of Photo-Optical Instrumentation Engineers (SPIE) Conference Series},
       volume = {9143},
        month = aug,
          eid = {914320},
        pages = {914320},
          doi = {10.1117/12.2063489},
archivePrefix = {arXiv},
       eprint = {1406.0151},
 primaryClass = {astro-ph.EP},
       adsurl = {https://ui.adsabs.harvard.edu/abs/2014SPIE.9143E..20R}
}

@Article{Gaia_mission2016,
  author        = {{Gaia Collaboration} and {Prusti}, T. and {de Bruijne}, J.~H.~J. and {Brown}, A.~G.~A. and {Vallenari}, A. and {Babusiaux}, C. and {Bailer-Jones}, C.~A.~L. and {Bastian}, U. and {Biermann}, M. and {Evans}, D.~W. and {Eyer}, L. and {Jansen}, F. and {Jordi}, C. and {Klioner}, S.~A. and {Lammers}, U. and {Lindegren}, L. and {Luri}, X. and {Mignard}, F. and {Milligan}, D.~J. and {Panem}, C. and {Poinsignon}, V. and {Pourbaix}, D. and {Randich}, S. and {Sarri}, G. and {Sartoretti}, P. and {Siddiqui}, H.~I. and {Soubiran}, C. and {Valette}, V. and {van Leeuwen}, F. and {Walton}, N.~A. and {Aerts}, C. and {Arenou}, F. and {Cropper}, M. and {Drimmel}, R. and {H{\o}g}, E. and {Katz}, D. and {Lattanzi}, M.~G. and {O'Mullane}, W. and {Grebel}, E.~K. and {Holland}, A.~D. and {Huc}, C. and {Passot}, X. and {Bramante}, L. and {Cacciari}, C. and {Casta{\~n}eda}, J. and {Chaoul}, L. and {Cheek}, N. and {De Angeli}, F. and {Fabricius}, C. and {Guerra}, R. and {Hern{\'a}ndez}, J. and {Jean-Antoine-Piccolo}, A. and {Masana}, E. and {Messineo}, R. and {Mowlavi}, N. and {Nienartowicz}, K. and {Ord{\'o}{\~n}ez-Blanco}, D. and {Panuzzo}, P. and {Portell}, J. and {Richards}, P.~J. and {Riello}, M. and {Seabroke}, G.~M. and {Tanga}, P. and {Th{\'e}venin}, F. and {Torra}, J. and {Els}, S.~G. and {Gracia-Abril}, G. and {Comoretto}, G. and {Garcia-Reinaldos}, M. and {Lock}, T. and {Mercier}, E. and {Altmann}, M. and {Andrae}, R. and {Astraatmadja}, T.~L. and {Bellas-Velidis}, I. and {Benson}, K. and {Berthier}, J. and {Blomme}, R. and {Busso}, G. and {Carry}, B. and {Cellino}, A. and {Clementini}, G. and {Cowell}, S. and {Creevey}, O. and {Cuypers}, J. and {Davidson}, M. and {De Ridder}, J. and {de Torres}, A. and {Delchambre}, L. and {Dell'Oro}, A. and {Ducourant}, C. and {Fr{\'e}mat}, Y. and {Garc{\'\i}a-Torres}, M. and {Gosset}, E. and {Halbwachs}, J. -L. and {Hambly}, N.~C. and {Harrison}, D.~L. and {Hauser}, M. and {Hestroffer}, D. and {Hodgkin}, S.~T. and {Huckle}, H.~E. and {Hutton}, A. and {Jasniewicz}, G. and {Jordan}, S. and {Kontizas}, M. and {Korn}, A.~J. and {Lanzafame}, A.~C. and {Manteiga}, M. and {Moitinho}, A. and {Muinonen}, K. and {Osinde}, J. and {Pancino}, E. and {Pauwels}, T. and {Petit}, J. -M. and {Recio-Blanco}, A. and {Robin}, A.~C. and {Sarro}, L.~M. and {Siopis}, C. and {Smith}, M. and {Smith}, K.~W. and {Sozzetti}, A. and {Thuillot}, W. and {van Reeven}, W. and {Viala}, Y. and {Abbas}, U. and {Abreu Aramburu}, A. and {Accart}, S. and {Aguado}, J.~J. and {Allan}, P.~M. and {Allasia}, W. and {Altavilla}, G. and {{\'A}lvarez}, M.~A. and {Alves}, J. and {Anderson}, R.~I. and {Andrei}, A.~H. and {Anglada Varela}, E. and {Antiche}, E. and {Antoja}, T. and {Ant{\'o}n}, S. and {Arcay}, B. and {Atzei}, A. and {Ayache}, L. and {Bach}, N. and {Baker}, S.~G. and {Balaguer-N{\'u}{\~n}ez}, L. and {Barache}, C. and {Barata}, C. and {Barbier}, A. and {Barblan}, F. and {Baroni}, M. and {Barrado y Navascu{\'e}s}, D. and {Barros}, M. and {Barstow}, M.~A. and {Becciani}, U. and {Bellazzini}, M. and {Bellei}, G. and {Bello Garc{\'\i}a}, A. and {Belokurov}, V. and {Bendjoya}, P. and {Berihuete}, A. and {Bianchi}, L. and {Bienaym{\'e}}, O. and {Billebaud}, F. and {Blagorodnova}, N. and {Blanco-Cuaresma}, S. and {Boch}, T. and {Bombrun}, A. and {Borrachero}, R. and {Bouquillon}, S. and {Bourda}, G. and {Bouy}, H. and {Bragaglia}, A. and {Breddels}, M.~A. and {Brouillet}, N. and {Br{\"u}semeister}, T. and {Bucciarelli}, B. and {Budnik}, F. and {Burgess}, P. and {Burgon}, R. and {Burlacu}, A. and {Busonero}, D. and {Buzzi}, R. and {Caffau}, E. and {Cambras}, J. and {Campbell}, H. and {Cancelliere}, R. and {Cantat-Gaudin}, T. and {Carlucci}, T. and {Carrasco}, J.~M. and {Castellani}, M. and {Charlot}, P. and {Charnas}, J. and {Charvet}, P. and {Chassat}, F. and {Chiavassa}, A. and {Clotet}, M. and {Cocozza}, G. and {Collins}, R.~S. and {Collins}, P. and {Costigan}, G. and {Crifo}, F. and {Cross}, N.~J.~G. and {Crosta}, M. and {Crowley}, C. and {Dafonte}, C. and {Damerdji}, Y. and {Dapergolas}, A. and {David}, P. and {David}, M. and {De Cat}, P. and {de Felice}, F. and {de Laverny}, P. and {De Luise}, F. and {De March}, R. and {de Martino}, D. and {de Souza}, R. and {Debosscher}, J. and {del Pozo}, E. and {Delbo}, M. and {Delgado}, A. and {Delgado}, H.~E. and {di Marco}, F. and {Di Matteo}, P. and {Diakite}, S. and {Distefano}, E. and {Dolding}, C. and {Dos Anjos}, S. and {Drazinos}, P. and {Dur{\'a}n}, J. and {Dzigan}, Y. and {Ecale}, E. and {Edvardsson}, B. and {Enke}, H. and {Erdmann}, M. and {Escolar}, D. and {Espina}, M. and {Evans}, N.~W. and {Eynard Bontemps}, G. and {Fabre}, C. and {Fabrizio}, M. and {Faigler}, S. and {Falc{\~a}o}, A.~J. and {Farr{\`a}s Casas}, M. and {Faye}, F. and {Federici}, L. and {Fedorets}, G. and {Fern{\'a}ndez-Hern{\'a}ndez}, J. and {Fernique}, P. and {Fienga}, A. and {Figueras}, F. and {Filippi}, F. and {Findeisen}, K. and {Fonti}, A. and {Fouesneau}, M. and {Fraile}, E. and {Fraser}, M. and {Fuchs}, J. and {Furnell}, R. and {Gai}, M. and {Galleti}, S. and {Galluccio}, L. and {Garabato}, D. and {Garc{\'\i}a-Sedano}, F. and {Gar{\'e}}, P. and {Garofalo}, A. and {Garralda}, N. and {Gavras}, P. and {Gerssen}, J. and {Geyer}, R. and {Gilmore}, G. and {Girona}, S. and {Giuffrida}, G. and {Gomes}, M. and {Gonz{\'a}lez-Marcos}, A. and {Gonz{\'a}lez-N{\'u}{\~n}ez}, J. and {Gonz{\'a}lez-Vidal}, J.~J. and {Granvik}, M. and {Guerrier}, A. and {Guillout}, P. and {Guiraud}, J. and {G{\'u}rpide}, A. and {Guti{\'e}rrez-S{\'a}nchez}, R. and {Guy}, L.~P. and {Haigron}, R. and {Hatzidimitriou}, D. and {Haywood}, M. and {Heiter}, U. and {Helmi}, A. and {Hobbs}, D. and {Hofmann}, W. and {Holl}, B. and {Holland}, G. and {Hunt}, J.~A.~S. and {Hypki}, A. and {Icardi}, V. and {Irwin}, M. and {Jevardat de Fombelle}, G. and {Jofr{\'e}}, P. and {Jonker}, P.~G. and {Jorissen}, A. and {Julbe}, F. and {Karampelas}, A. and {Kochoska}, A. and {Kohley}, R. and {Kolenberg}, K. and {Kontizas}, E. and {Koposov}, S.~E. and {Kordopatis}, G. and {Koubsky}, P. and {Kowalczyk}, A. and {Krone-Martins}, A. and {Kudryashova}, M. and {Kull}, I. and {Bachchan}, R.~K. and {Lacoste-Seris}, F. and {Lanza}, A.~F. and {Lavigne}, J. -B. and {Le Poncin-Lafitte}, C. and {Lebreton}, Y. and {Lebzelter}, T. and {Leccia}, S. and {Leclerc}, N. and {Lecoeur-Taibi}, I. and {Lemaitre}, V. and {Lenhardt}, H. and {Leroux}, F. and {Liao}, S. and {Licata}, E. and {Lindstr{\o}m}, H.~E.~P. and {Lister}, T.~A. and {Livanou}, E. and {Lobel}, A. and {L{\"o}ffler}, W. and {L{\'o}pez}, M. and {Lopez-Lozano}, A. and {Lorenz}, D. and {Loureiro}, T. and {MacDonald}, I. and {Magalh{\~a}es Fernandes}, T. and {Managau}, S. and {Mann}, R.~G. and {Mantelet}, G. and {Marchal}, O. and {Marchant}, J.~M. and {Marconi}, M. and {Marie}, J. and {Marinoni}, S. and {Marrese}, P.~M. and {Marschalk{\'o}}, G. and {Marshall}, D.~J. and {Mart{\'\i}n-Fleitas}, J.~M. and {Martino}, M. and {Mary}, N. and {Matijevi{\v{c}}}, G. and {Mazeh}, T. and {McMillan}, P.~J. and {Messina}, S. and {Mestre}, A. and {Michalik}, D. and {Millar}, N.~R. and {Miranda}, B.~M.~H. and {Molina}, D. and {Molinaro}, R. and {Molinaro}, M. and {Moln{\'a}r}, L. and {Moniez}, M. and {Montegriffo}, P. and {Monteiro}, D. and {Mor}, R. and {Mora}, A. and {Morbidelli}, R. and {Morel}, T. and {Morgenthaler}, S. and {Morley}, T. and {Morris}, D. and {Mulone}, A.~F. and {Muraveva}, T. and {Musella}, I. and {Narbonne}, J. and {Nelemans}, G. and {Nicastro}, L. and {Noval}, L. and {Ord{\'e}novic}, C. and {Ordieres-Mer{\'e}}, J. and {Osborne}, P. and {Pagani}, C. and {Pagano}, I. and {Pailler}, F. and {Palacin}, H. and {Palaversa}, L. and {Parsons}, P. and {Paulsen}, T. and {Pecoraro}, M. and {Pedrosa}, R. and {Pentik{\"a}inen}, H. and {Pereira}, J. and {Pichon}, B. and {Piersimoni}, A.~M. and {Pineau}, F. -X. and {Plachy}, E. and {Plum}, G. and {Poujoulet}, E. and {Pr{\v{s}}a}, A. and {Pulone}, L. and {Ragaini}, S. and {Rago}, S. and {Rambaux}, N. and {Ramos-Lerate}, M. and {Ranalli}, P. and {Rauw}, G. and {Read}, A. and {Regibo}, S. and {Renk}, F. and {Reyl{\'e}}, C. and {Ribeiro}, R.~A. and {Rimoldini}, L. and {Ripepi}, V. and {Riva}, A. and {Rixon}, G. and {Roelens}, M. and {Romero-G{\'o}mez}, M. and {Rowell}, N. and {Royer}, F. and {Rudolph}, A. and {Ruiz-Dern}, L. and {Sadowski}, G. and {Sagrist{\`a} Sell{\'e}s}, T. and {Sahlmann}, J. and {Salgado}, J. and {Salguero}, E. and {Sarasso}, M. and {Savietto}, H. and {Schnorhk}, A. and {Schultheis}, M. and {Sciacca}, E. and {Segol}, M. and {Segovia}, J.~C. and {Segransan}, D. and {Serpell}, E. and {Shih}, I. -C. and {Smareglia}, R. and {Smart}, R.~L. and {Smith}, C. and {Solano}, E. and {Solitro}, F. and {Sordo}, R. and {Soria Nieto}, S. and {Souchay}, J. and {Spagna}, A. and {Spoto}, F. and {Stampa}, U. and {Steele}, I.~A. and {Steidelm{\"u}ller}, H. and {Stephenson}, C.~A. and {Stoev}, H. and {Suess}, F.~F. and {S{\"u}veges}, M. and {Surdej}, J. and {Szabados}, L. and {Szegedi-Elek}, E. and {Tapiador}, D. and {Taris}, F. and {Tauran}, G. and {Taylor}, M.~B. and {Teixeira}, R. and {Terrett}, D. and {Tingley}, B. and {Trager}, S.~C. and {Turon}, C. and {Ulla}, A. and {Utrilla}, E. and {Valentini}, G. and {van Elteren}, A. and {Van Hemelryck}, E. and {van Leeuwen}, M. and {Varadi}, M. and {Vecchiato}, A. and {Veljanoski}, J. and {Via}, T. and {Vicente}, D. and {Vogt}, S. and {Voss}, H. and {Votruba}, V. and {Voutsinas}, S. and {Walmsley}, G. and {Weiler}, M. and {Weingrill}, K. and {Werner}, D. and {Wevers}, T. and {Whitehead}, G. and {Wyrzykowski}, {\L}. and {Yoldas}, A. and {{\v{Z}}erjal}, M. and {Zucker}, S. and {Zurbach}, C. and {Zwitter}, T. and {Alecu}, A. and {Allen}, M. and {Allende Prieto}, C. and {Amorim}, A. and {Anglada-Escud{\'e}}, G. and {Arsenijevic}, V. and {Azaz}, S. and {Balm}, P. and {Beck}, M. and {Bernstein}, H. -H. and {Bigot}, L. and {Bijaoui}, A. and {Blasco}, C. and {Bonfigli}, M. and {Bono}, G. and {Boudreault}, S. and {Bressan}, A. and {Brown}, S. and {Brunet}, P. -M. and {Bunclark}, P. and {Buonanno}, R. and {Butkevich}, A.~G. and {Carret}, C. and {Carrion}, C. and {Chemin}, L. and {Ch{\'e}reau}, F. and {Corcione}, L. and {Darmigny}, E. and {de Boer}, K.~S. and {de Teodoro}, P. and {de Zeeuw}, P.~T. and {Delle Luche}, C. and {Domingues}, C.~D. and {Dubath}, P. and {Fodor}, F. and {Fr{\'e}zouls}, B. and {Fries}, A. and {Fustes}, D. and {Fyfe}, D. and {Gallardo}, E. and {Gallegos}, J. and {Gardiol}, D. and {Gebran}, M. and {Gomboc}, A. and {G{\'o}mez}, A. and {Grux}, E. and {Gueguen}, A. and {Heyrovsky}, A. and {Hoar}, J. and {Iannicola}, G. and {Isasi Parache}, Y. and {Janotto}, A. -M. and {Joliet}, E. and {Jonckheere}, A. and {Keil}, R. and {Kim}, D. -W. and {Klagyivik}, P. and {Klar}, J. and {Knude}, J. and {Kochukhov}, O. and {Kolka}, I. and {Kos}, J. and {Kutka}, A. and {Lainey}, V. and {LeBouquin}, D. and {Liu}, C. and {Loreggia}, D. and {Makarov}, V.~V. and {Marseille}, M.~G. and {Martayan}, C. and {Martinez-Rubi}, O. and {Massart}, B. and {Meynadier}, F. and {Mignot}, S. and {Munari}, U. and {Nguyen}, A. -T. and {Nordlander}, T. and {Ocvirk}, P. and {O'Flaherty}, K.~S. and {Olias Sanz}, A. and {Ortiz}, P. and {Osorio}, J. and {Oszkiewicz}, D. and {Ouzounis}, A. and {Palmer}, M. and {Park}, P. and {Pasquato}, E. and {Peltzer}, C. and {Peralta}, J. and {P{\'e}turaud}, F. and {Pieniluoma}, T. and {Pigozzi}, E. and {Poels}, J. and {Prat}, G. and {Prod'homme}, T. and {Raison}, F. and {Rebordao}, J.~M. and {Risquez}, D. and {Rocca-Volmerange}, B. and {Rosen}, S. and {Ruiz-Fuertes}, M.~I. and {Russo}, F. and {Sembay}, S. and {Serraller Vizcaino}, I. and {Short}, A. and {Siebert}, A. and {Silva}, H. and {Sinachopoulos}, D. and {Slezak}, E. and {Soffel}, M. and {Sosnowska}, D. and {Strai{\v{z}}ys}, V. and {ter Linden}, M. and {Terrell}, D. and {Theil}, S. and {Tiede}, C. and {Troisi}, L. and {Tsalmantza}, P. and {Tur}, D. and {Vaccari}, M. and {Vachier}, F. and {Valles}, P. and {Van Hamme}, W. and {Veltz}, L. and {Virtanen}, J. and {Wallut}, J. -M. and {Wichmann}, R. and {Wilkinson}, M.~I. and {Ziaeepour}, H. and {Zschocke}, S.},
  title         = {{The Gaia mission}},
  doi           = {10.1051/0004-6361/201629272},
  eid           = {A1},
  eprint        = {1609.04153},
  pages         = {A1},
  volume        = {595},
  adsurl        = {https://ui.adsabs.harvard.edu/abs/2016A&A...595A...1G},
  archiveprefix = {arXiv},
  journal       = {\aap},
  month         = nov,
  primaryclass  = {astro-ph.IM},
  year          = {2016},
}

@Article{GaiaEDR3Validation,
  author        = {{Fabricius}, C. and {Luri}, X. and {Arenou}, F. and {Babusiaux}, C. and {Helmi}, A. and {Muraveva}, T. and {Reyl{\'e}}, C. and {Spoto}, F. and {Vallenari}, A. and {Antoja}, T. and {Balbinot}, E. and {Barache}, C. and {Bauchet}, N. and {Bragaglia}, A. and {Busonero}, D. and {Cantat-Gaudin}, T. and {Carrasco}, J.~M. and {Diakit{\'e}}, S. and {Fabrizio}, M. and {Figueras}, F. and {Garcia-Gutierrez}, A. and {Garofalo}, A. and {Jordi}, C. and {Kervella}, P. and {Khanna}, S. and {Leclerc}, N. and {Licata}, E. and {Lambert}, S. and {Marrese}, P.~M. and {Masip}, A. and {Ramos}, P. and {Robichon}, N. and {Robin}, A.~C. and {Romero-G{\'o}mez}, M. and {Rubele}, S. and {Weiler}, M.},
  title         = {{Gaia Early Data Release 3. Catalogue validation}},
  doi           = {10.1051/0004-6361/202039834},
  eid           = {A5},
  eprint        = {2012.06242},
  pages         = {A5},
  volume        = {649},
  adsurl        = {https://ui.adsabs.harvard.edu/abs/2021A&A...649A...5F},
  archiveprefix = {arXiv},
  journal       = {\aap},
  month         = may,
  primaryclass  = {astro-ph.GA},
  year          = {2021},
}

@ARTICLE{lin21a,
       author = {{Lindegren}, L. and {Bastian}, U. and {Biermann}, M. and {Bombrun}, A. and {de Torres}, A. and {Gerlach}, E. and {Geyer}, R. and {Hern{\'a}ndez}, J. and {Hilger}, T. and {Hobbs}, D. and {Klioner}, S.~A. and {Lammers}, U. and {McMillan}, P.~J. and {Ramos-Lerate}, M. and {Steidelm{\"u}ller}, H. and {Stephenson}, C.~A. and {van Leeuwen}, F.},
        title = "{Gaia Early Data Release 3. Parallax bias versus magnitude, colour, and position}",
      journal = {\aap},
         year = 2021,
        month = may,
       volume = {649},
          eid = {A4},
        pages = {A4},
          doi = {10.1051/0004-6361/202039653},
archivePrefix = {arXiv},
       eprint = {2012.01742},
 primaryClass = {astro-ph.IM},
       adsurl = {https://ui.adsabs.harvard.edu/abs/2021A&A...649A...4L}
}

@ARTICLE{lin21b,
       author = {{Lindegren}, L. and {Klioner}, S.~A. and {Hern{\'a}ndez}, J. and {Bombrun}, A. and {Ramos-Lerate}, M. and {Steidelm{\"u}ller}, H. and {Bastian}, U. and {Biermann}, M. and {de Torres}, A. and {Gerlach}, E. and {Geyer}, R. and {Hilger}, T. and {Hobbs}, D. and {Lammers}, U. and {McMillan}, P.~J. and {Stephenson}, C.~A. and {Casta{\~n}eda}, J. and {Davidson}, M. and {Fabricius}, C. and {Gracia-Abril}, G. and {Portell}, J. and {Rowell}, N. and {Teyssier}, D. and {Torra}, F. and {Bartolom{\'e}}, S. and {Clotet}, M. and {Garralda}, N. and {Gonz{\'a}lez-Vidal}, J.~J. and {Torra}, J. and {Abbas}, U. and {Altmann}, M. and {Anglada Varela}, E. and {Balaguer-N{\'u}{\~n}ez}, L. and {Balog}, Z. and {Barache}, C. and {Becciani}, U. and {Bernet}, M. and {Bertone}, S. and {Bianchi}, L. and {Bouquillon}, S. and {Brown}, A.~G.~A. and {Bucciarelli}, B. and {Busonero}, D. and {Butkevich}, A.~G. and {Buzzi}, R. and {Cancelliere}, R. and {Carlucci}, T. and {Charlot}, P. and {Cioni}, M. -R.~L. and {Crosta}, M. and {Crowley}, C. and {del Peloso}, E.~F. and {del Pozo}, E. and {Drimmel}, R. and {Esquej}, P. and {Fienga}, A. and {Fraile}, E. and {Gai}, M. and {Garcia-Reinaldos}, M. and {Guerra}, R. and {Hambly}, N.~C. and {Hauser}, M. and {Jan{\ss}en}, K. and {Jordan}, S. and {Kostrzewa-Rutkowska}, Z. and {Lattanzi}, M.~G. and {Liao}, S. and {Licata}, E. and {Lister}, T.~A. and {L{\"o}ffler}, W. and {Marchant}, J.~M. and {Masip}, A. and {Mignard}, F. and {Mints}, A. and {Molina}, D. and {Mora}, A. and {Morbidelli}, R. and {Murphy}, C.~P. and {Pagani}, C. and {Panuzzo}, P. and {Pe{\~n}alosa Esteller}, X. and {Poggio}, E. and {Re Fiorentin}, P. and {Riva}, A. and {Sagrist{\`a} Sell{\'e}s}, A. and {Sanchez Gimenez}, V. and {Sarasso}, M. and {Sciacca}, E. and {Siddiqui}, H.~I. and {Smart}, R.~L. and {Souami}, D. and {Spagna}, A. and {Steele}, I.~A. and {Taris}, F. and {Utrilla}, E. and {van Reeven}, W. and {Vecchiato}, A.},
        title = "{Gaia Early Data Release 3. The astrometric solution}",
      journal = {\aap},
         year = 2021,
        month = may,
       volume = {649},
          eid = {A2},
        pages = {A2},
          doi = {10.1051/0004-6361/202039709},
archivePrefix = {arXiv},
       eprint = {2012.03380},
 primaryClass = {astro-ph.IM},
       adsurl = {https://ui.adsabs.harvard.edu/abs/2021A&A...649A...2L}
}

@Article{EDR3photometry,
  author        = {{Riello}, M. and {De Angeli}, F. and {Evans}, D.~W. and {Montegriffo}, P. and {Carrasco}, J.~M. and {Busso}, G. and {Palaversa}, L. and {Burgess}, P.~W. and {Diener}, C. and {Davidson}, M. and {Rowell}, N. and {Fabricius}, C. and {Jordi}, C. and {Bellazzini}, M. and {Pancino}, E. and {Harrison}, D.~L. and {Cacciari}, C. and {van Leeuwen}, F. and {Hambly}, N.~C. and {Hodgkin}, S.~T. and {Osborne}, P.~J. and {Altavilla}, G. and {Barstow}, M.~A. and {Brown}, A.~G.~A. and {Castellani}, M. and {Cowell}, S. and {De Luise}, F. and {Gilmore}, G. and {Giuffrida}, G. and {Hidalgo}, S. and {Holland}, G. and {Marinoni}, S. and {Pagani}, C. and {Piersimoni}, A.~M. and {Pulone}, L. and {Ragaini}, S. and {Rainer}, M. and {Richards}, P.~J. and {Sanna}, N. and {Walton}, N.~A. and {Weiler}, M. and {Yoldas}, A.},
  title         = {{Gaia Early Data Release 3. Photometric content and validation}},
  doi           = {10.1051/0004-6361/202039587},
  eid           = {A3},
  eprint        = {2012.01916},
  pages         = {A3},
  volume        = {649},
  adsurl        = {https://ui.adsabs.harvard.edu/abs/2021A&A...649A...3R},
  archiveprefix = {arXiv},
  journal       = {\aap},
  month         = may,
  primaryclass  = {astro-ph.IM},
  year          = {2021},
}

@ARTICLE{row21,
       author = {{Rowell}, N. and {Davidson}, M. and {Lindegren}, L. and {van Leeuwen}, F. and {Casta{\~n}eda}, J. and {Fabricius}, C. and {Bastian}, U. and {Hambly}, N.~C. and {Hern{\'a}ndez}, J. and {Bombrun}, A. and {Evans}, D.~W. and {De Angeli}, F. and {Riello}, M. and {Busonero}, D. and {Crowley}, C. and {Mora}, A. and {Lammers}, U. and {Gracia}, G. and {Portell}, J. and {Biermann}, M. and {Brown}, A.~G.~A.},
        title = "{Gaia Early Data Release 3. Modelling and calibration of Gaia's point and line spread functions}",
      journal = {\aap},
         year = 2021,
        month = may,
       volume = {649},
          eid = {A11},
        pages = {A11},
          doi = {10.1051/0004-6361/202039448},
archivePrefix = {arXiv},
       eprint = {2012.02069},
 primaryClass = {astro-ph.IM},
       adsurl = {https://ui.adsabs.harvard.edu/abs/2021A&A...649A..11R}
}

@ARTICLE{tor21,
       author = {{Torra}, F. and {Casta{\~n}eda}, J. and {Fabricius}, C. and {Lindegren}, L. and {Clotet}, M. and {Gonz{\'a}lez-Vidal}, J.~J. and {Bartolom{\'e}}, S. and {Bastian}, U. and {Bernet}, M. and {Biermann}, M. and {Garralda}, N. and {G{\'u}rpide}, A. and {Lammers}, U. and {Portell}, J. and {Torra}, J.},
        title = "{Gaia Early Data Release 3. Building the Gaia DR3 source list - Cross-match of Gaia observations}",
      journal = {\aap},
         year = 2021,
        month = may,
       volume = {649},
          eid = {A10},
        pages = {A10},
          doi = {10.1051/0004-6361/202039637},
archivePrefix = {arXiv},
       eprint = {2012.06420},
 primaryClass = {astro-ph.IM},
       adsurl = {https://ui.adsabs.harvard.edu/abs/2021A&A...649A..10T}
}

@Article{Ziegler2020,
  author        = {{Ziegler}, Carl and {Tokovinin}, Andrei and {Brice{\~n}o}, C{\'e}sar and {Mang}, James and {Law}, Nicholas and {Mann}, Andrew W.},
  title         = {{SOAR TESS Survey. I. Sculpting of TESS Planetary Systems by Stellar Companions}},
  doi           = {10.3847/1538-3881/ab55e9},
  eid           = {19},
  eprint        = {1908.10871},
  number        = {1},
  pages         = {19},
  volume        = {159},
  adsurl        = {https://ui.adsabs.harvard.edu/abs/2020AJ....159...19Z},
  archiveprefix = {arXiv},
  journal       = {\aj},
  month         = jan,
  primaryclass  = {astro-ph.EP},
  year          = {2020},
}

@Article{GaiaDR3Frame,
  author        = {{Gaia Collaboration} and {Klioner}, S.~A. and {Lindegren}, L. and {Mignard}, F. and {Hern{\'a}ndez}, J. and {Ramos-Lerate}, M. and {Bastian}, U. and {Biermann}, M. and {Bombrun}, A. and {de Torres}, A. and {Gerlach}, E. and {Geyer}, R. and {Hilger}, T. and {Hobbs}, D. and {Lammers}, U.~L. and {McMillan}, P.~J. and {Steidelm{\"u}ller}, H. and {Teyssier}, D. and {Raiteri}, C.~M. and {Bartolom{\'e}}, S. and {Bernet}, M. and {Casta{\~n}eda}, J. and {Clotet}, M. and {Davidson}, M. and {Fabricius}, C. and {Garralda Torres}, N. and {Gonz{\'a}lez-Vidal}, J.~J. and {Portell}, J. and {Rowell}, N. and {Torra}, F. and {Torra}, J. and {Brown}, A.~G.~A. and {Vallenari}, A. and {Prusti}, T. and {de Bruijne}, J.~H.~J. and {Arenou}, F. and {Babusiaux}, C. and {Creevey}, O.~L. and {Ducourant}, C. and {Evans}, D.~W. and {Eyer}, L. and {Guerra}, R. and {Hutton}, A. and {Jordi}, C. and {Luri}, X. and {Panem}, C. and {Pourbaix}, D. and {Randich}, S. and {Sartoretti}, P. and {Soubiran}, C. and {Tanga}, P. and {Walton}, N.~A. and {Bailer-Jones}, C.~A.~L. and {Drimmel}, R. and {Jansen}, F. and {Katz}, D. and {Lattanzi}, M.~G. and {van Leeuwen}, F. and {Bakker}, J. and {Cacciari}, C. and {De Angeli}, F. and {Fouesneau}, M. and {Fr{\'e}mat}, Y. and {Galluccio}, L. and {Guerrier}, A. and {Heiter}, U. and {Masana}, E. and {Messineo}, R. and {Mowlavi}, N. and {Nicolas}, C. and {Nienartowicz}, K. and {Pailler}, F. and {Panuzzo}, P. and {Riclet}, F. and {Roux}, W. and {Seabroke}, G.~M. and {Sordo}, R. and {Th{\'e}venin}, F. and {Gracia-Abril}, G. and {Altmann}, M. and {Andrae}, R. and {Audard}, M. and {Bellas-Velidis}, I. and {Benson}, K. and {Berthier}, J. and {Blomme}, R. and {Burgess}, P.~W. and {Busonero}, D. and {Busso}, G. and {C{\'a}novas}, H. and {Carry}, B. and {Cellino}, A. and {Cheek}, N. and {Clementini}, G. and {Damerdji}, Y. and {de Teodoro}, P. and {Nu{\~n}ez Campos}, M. and {Delchambre}, L. and {Dell'Oro}, A. and {Esquej}, P. and {Fern{\'a}ndez-Hern{\'a}ndez}, J. and {Fraile}, E. and {Garabato}, D. and {Garc{\'\i}a-Lario}, P. and {Gosset}, E. and {Haigron}, R. and {Halbwachs}, J. -L. and {Hambly}, N.~C. and {Harrison}, D.~L. and {Hestroffer}, D. and {Hodgkin}, S.~T. and {Holl}, B. and {Jan{\ss}en}, K. and {Jevardat de Fombelle}, G. and {Jordan}, S. and {Krone-Martins}, A. and {Lanzafame}, A.~C. and {L{\"o}ffler}, W. and {Marchal}, O. and {Marrese}, P.~M. and {Moitinho}, A. and {Muinonen}, K. and {Osborne}, P. and {Pancino}, E. and {Pauwels}, T. and {Recio-Blanco}, A. and {Reyl{\'e}}, C. and {Riello}, M. and {Rimoldini}, L. and {Roegiers}, T. and {Rybizki}, J. and {Sarro}, L.~M. and {Siopis}, C. and {Smith}, M. and {Sozzetti}, A. and {Utrilla}, E. and {van Leeuwen}, M. and {Abbas}, U. and {{\'A}brah{\'a}m}, P. and {Abreu Aramburu}, A. and {Aerts}, C. and {Aguado}, J.~J. and {Ajaj}, M. and {Aldea-Montero}, F. and {Altavilla}, G. and {{\'A}lvarez}, M.~A. and {Alves}, J. and {Anderson}, R.~I. and {Anglada Varela}, E. and {Antoja}, T. and {Baines}, D. and {Baker}, S.~G. and {Balaguer-N{\'u}{\~n}ez}, L. and {Balbinot}, E. and {Balog}, Z. and {Barache}, C. and {Barbato}, D. and {Barros}, M. and {Barstow}, M.~A. and {Bassilana}, J. -L. and {Bauchet}, N. and {Becciani}, U. and {Bellazzini}, M. and {Berihuete}, A. and {Bertone}, S. and {Bianchi}, L. and {Binnenfeld}, A. and {Blanco-Cuaresma}, S. and {Boch}, T. and {Bossini}, D. and {Bouquillon}, S. and {Bragaglia}, A. and {Bramante}, L. and {Breedt}, E. and {Bressan}, A. and {Brouillet}, N. and {Brugaletta}, E. and {Bucciarelli}, B. and {Burlacu}, A. and {Butkevich}, A.~G. and {Buzzi}, R. and {Caffau}, E. and {Cancelliere}, R. and {Cantat-Gaudin}, T. and {Carballo}, R. and {Carlucci}, T. and {Carnerero}, M.~I. and {Carrasco}, J.~M. and {Casamiquela}, L. and {Castellani}, M. and {Castro-Ginard}, A. and {Chaoul}, L. and {Charlot}, P. and {Chemin}, L. and {Chiaramida}, V. and {Chiavassa}, A. and {Chornay}, N. and {Comoretto}, G. and {Contursi}, G. and {Cooper}, W.~J. and {Cornez}, T. and {Cowell}, S. and {Crifo}, F. and {Cropper}, M. and {Crosta}, M. and {Crowley}, C. and {Dafonte}, C. and {Dapergolas}, A. and {David}, P. and {de Laverny}, P. and {De Luise}, F. and {De March}, R. and {De Ridder}, J. and {de Souza}, R. and {del Peloso}, E.~F. and {del Pozo}, E. and {Delbo}, M. and {Delgado}, A. and {Delisle}, J. -B. and {Demouchy}, C. and {Dharmawardena}, T.~E. and {Diakite}, S. and {Diener}, C. and {Distefano}, E. and {Dolding}, C. and {Enke}, H. and {Fabre}, C. and {Fabrizio}, M. and {Faigler}, S. and {Fedorets}, G. and {Fernique}, P. and {Fienga}, A. and {Figueras}, F. and {Fournier}, Y. and {Fouron}, C. and {Fragkoudi}, F. and {Gai}, M. and {Garcia-Gutierrez}, A. and {Garcia-Reinaldos}, M. and {Garc{\'\i}a-Torres}, M. and {Garofalo}, A. and {Gavel}, A. and {Gavras}, P. and {Giacobbe}, P. and {Gilmore}, G. and {Girona}, S. and {Giuffrida}, G. and {Gomel}, R. and {Gomez}, A. and {Gonz{\'a}lez-N{\'u}{\~n}ez}, J. and {Gonz{\'a}lez-Santamar{\'\i}a}, I. and {Granvik}, M. and {Guillout}, P. and {Guiraud}, J. and {Guti{\'e}rrez-S{\'a}nchez}, R. and {Guy}, L.~P. and {Hatzidimitriou}, D. and {Hauser}, M. and {Haywood}, M. and {Helmer}, A. and {Helmi}, A. and {Sarmiento}, M.~H. and {Hidalgo}, S.~L. and {H{\l}adczuk}, N. and {Holland}, G. and {Huckle}, H.~E. and {Jardine}, K. and {Jasniewicz}, G. and {Jean-Antoine Piccolo}, A. and {Jim{\'e}nez-Arranz}, {\'O}. and {Juaristi Campillo}, J. and {Julbe}, F. and {Karbevska}, L. and {Kervella}, P. and {Khanna}, S. and {Kordopatis}, G. and {Korn}, A.~J. and {K{\'o}sp{\'a}l}, {\'A}. and {Kostrzewa-Rutkowska}, Z. and {Kruszy{\'n}ska}, K. and {Kun}, M. and {Laizeau}, P. and {Lambert}, S. and {Lanza}, A.~F. and {Lasne}, Y. and {Le Campion}, J. -F. and {Lebreton}, Y. and {Lebzelter}, T. and {Leccia}, S. and {Leclerc}, N. and {Lecoeur-Taibi}, I. and {Liao}, S. and {Licata}, E.~L. and {Lindstr{\o}m}, H.~E.~P. and {Lister}, T.~A. and {Livanou}, E. and {Lobel}, A. and {Lorca}, A. and {Loup}, C. and {Madrero Pardo}, P. and {Magdaleno Romeo}, A. and {Managau}, S. and {Mann}, R.~G. and {Manteiga}, M. and {Marchant}, J.~M. and {Marconi}, M. and {Marcos}, J. and {Santos}, M.~M.~S. Marcos and {Mar{\'\i}n Pina}, D. and {Marinoni}, S. and {Marocco}, F. and {Marshall}, D.~J. and {Polo}, L. Martin and {Mart{\'\i}n-Fleitas}, J.~M. and {Marton}, G. and {Mary}, N. and {Masip}, A. and {Massari}, D. and {Mastrobuono-Battisti}, A. and {Mazeh}, T. and {Messina}, S. and {Michalik}, D. and {Millar}, N.~R. and {Mints}, A. and {Molina}, D. and {Molinaro}, R. and {Moln{\'a}r}, L. and {Monari}, G. and {Mongui{\'o}}, M. and {Montegriffo}, P. and {Montero}, A. and {Mor}, R. and {Mora}, A. and {Morbidelli}, R. and {Morel}, T. and {Morris}, D. and {Muraveva}, T. and {Murphy}, C.~P. and {Musella}, I. and {Nagy}, Z. and {Noval}, L. and {Oca{\~n}a}, F. and {Ogden}, A. and {Ordenovic}, C. and {Osinde}, J.~O. and {Pagani}, C. and {Pagano}, I. and {Palaversa}, L. and {Palicio}, P.~A. and {Pallas-Quintela}, L. and {Panahi}, A. and {Payne-Wardenaar}, S. and {Pe{\~n}alosa Esteller}, X. and {Penttil{\"a}}, A. and {Pichon}, B. and {Piersimoni}, A.~M. and {Pineau}, F. -X. and {Plachy}, E. and {Plum}, G. and {Poggio}, E. and {Pr{\v{s}}a}, A. and {Pulone}, L. and {Racero}, E. and {Ragaini}, S. and {Rainer}, M. and {Rambaux}, N. and {Ramos}, P. and {Re Fiorentin}, P. and {Regibo}, S. and {Richards}, P.~J. and {Diaz}, C. Rios and {Ripepi}, V. and {Riva}, A. and {Rix}, H. -W. and {Rixon}, G. and {Robichon}, N. and {Robin}, A.~C. and {Robin}, C. and {Roelens}, M. and {Rogues}, H.~R.~O. and {Rohrbasser}, L. and {Romero-G{\'o}mez}, M. and {Royer}, F. and {Ruz Mieres}, D. and {Rybicki}, K.~A. and {Sadowski}, G. and {S{\'a}ez N{\'u}{\~n}ez}, A. and {Sagrist{\`a} Sell{\'e}s}, A. and {Sahlmann}, J. and {Salguero}, E. and {Samaras}, N. and {Sanchez Gimenez}, V. and {Sanna}, N. and {Santove{\~n}a}, R. and {Sarasso}, M. and {Schultheis}, M. and {Sciacca}, E. and {Segol}, M. and {Segovia}, J.~C. and {S{\'e}gransan}, D. and {Semeux}, D. and {Shahaf}, S. and {Siddiqui}, H.~I. and {Siebert}, A. and {Siltala}, L. and {Silvelo}, A. and {Slezak}, E. and {Slezak}, I. and {Smart}, R.~L. and {Snaith}, O.~N. and {Solano}, E. and {Solitro}, F. and {Souami}, D. and {Souchay}, J. and {Spagna}, A. and {Spina}, L. and {Spoto}, F. and {Steele}, I.~A. and {Stephenson}, C.~A. and {S{\"u}veges}, M. and {Surdej}, J. and {Szabados}, L. and {Szegedi-Elek}, E. and {Taris}, F. and {Taylor}, M.~B. and {Teixeira}, R. and {Tolomei}, L. and {Tonello}, N. and {Torralba Elipe}, G. and {Trabucchi}, M. and {Tsounis}, A.~T. and {Turon}, C. and {Ulla}, A. and {Unger}, N. and {Vaillant}, M.~V. and {van Dillen}, E. and {van Reeven}, W. and {Vanel}, O. and {Vecchiato}, A. and {Viala}, Y. and {Vicente}, D. and {Voutsinas}, S. and {Weiler}, M. and {Wevers}, T. and {Wyrzykowski}, {\L}. and {Yoldas}, A. and {Yvard}, P. and {Zhao}, H. and {Zorec}, J. and {Zucker}, S. and {Zwitter}, T.},
  title         = {{Gaia Early Data Release 3. The celestial reference frame (Gaia-CRF3)}},
  doi           = {10.1051/0004-6361/202243483},
  eid           = {A148},
  eprint        = {2204.12574},
  pages         = {A148},
  volume        = {667},
  adsurl        = {https://ui.adsabs.harvard.edu/abs/2022A&A...667A.148G},
  archiveprefix = {arXiv},
  journal       = {\aap},
  month         = nov,
  primaryclass  = {astro-ph.IM},
  year          = {2022},
}

@Misc{GaiaEDR3XMatch,
  author       = {{Marrese}, P.~M. and {Marinoni}, S. and {Fabrizio}, M. and {Altavilla}, G.},
  title        = {{Gaia EDR3 documentation Chapter 9: Cross-match with external catalogues}},
  howpublished = {Gaia EDR3 documentation, European Space Agency; Gaia Data Processing and Analysis Consortium. Online at <A href=``https://gea.esac.esa.int/archive/documentation/GEDR3/index.html''>https://gea.esac.esa.int/archive/documentation/GEDR3/index.html</A>, id. 9},
  adsurl       = {https://ui.adsabs.harvard.edu/abs/2021gdr3.reptE...9M},
  eid          = {9},
  month        = mar,
  pages        = {9},
  year         = {2021},
}

@Misc{GaiaDR3XMatch,
  author       = {{Marrese}, P.~M. and {Marinoni}, S. and {Fabrizio}, M. and {Altavilla}, G.},
  title        = {{Gaia DR3 documentation Chapter 15: Cross-match with external catalogues}},
  howpublished = {Gaia DR3 documentation, European Space Agency; Gaia Data Processing and Analysis Consortium. Online at <A href=``https://gea.esac.esa.int/archive/documentation/GDR3/index.html''>https://gea.esac.esa.int/archive/documentation/GDR3/index.html</A>, id. 15},
  adsurl       = {https://ui.adsabs.harvard.edu/abs/2022gdr3.reptE..15M},
  eid          = {15},
  month        = jun,
  pages        = {15},
  year         = {2022},
}

@InProceedings{TOPCAT,
  author    = {{Taylor}, M.~B.},
  booktitle = {Astronomical Data Analysis Software and Systems XIV},
  title     = {{TOPCAT \& STIL: Starlink Table/VOTable Processing Software}},
  editor    = {{Shopbell}, P. and {Britton}, M. and {Ebert}, R.},
  pages     = {29},
  series    = {Astronomical Society of the Pacific Conference Series},
  volume    = {347},
  adsurl    = {https://ui.adsabs.harvard.edu/abs/2005ASPC..347...29T},
  month     = dec,
  year      = {2005},
}

@Article{AstropyI,
  author        = {{Astropy Collaboration} and {Robitaille}, T.~P. and {Tollerud}, E.~J. and {Greenfield}, P. and {Droettboom}, M. and {Bray}, E. and {Aldcroft}, T. and {Davis}, M. and {Ginsburg}, A. and {Price-Whelan}, A.~M. and {Kerzendorf}, W.~E. and {Conley}, A. and {Crighton}, N. and {Barbary}, K. and {Muna}, D. and {Ferguson}, H. and {Grollier}, F. and {Parikh}, M.~M. and {Nair}, P.~H. and {Unther}, H.~M. and {Deil}, C. and {Woillez}, J. and {Conseil}, S. and {Kramer}, R. and {Turner}, J.~E.~H. and {Singer}, L. and {Fox}, R. and {Weaver}, B.~A. and {Zabalza}, V. and {Edwards}, Z.~I. and {Azalee Bostroem}, K. and {Burke}, D.~J. and {Casey}, A.~R. and {Crawford}, S.~M. and {Dencheva}, N. and {Ely}, J. and {Jenness}, T. and {Labrie}, K. and {Lim}, P.~L. and {Pierfederici}, F. and {Pontzen}, A. and {Ptak}, A. and {Refsdal}, B. and {Servillat}, M. and {Streicher}, O.},
  title         = {{Astropy: A community Python package for astronomy}},
  doi           = {10.1051/0004-6361/201322068},
  eid           = {A33},
  eprint        = {1307.6212},
  pages         = {A33},
  volume        = {558},
  adsurl        = {http://adsabs.harvard.edu/abs/2013A%26A...558A..33A},
  archiveprefix = {arXiv},
  journal       = {\aap},
  month         = oct,
  primaryclass  = {astro-ph.IM},
  year          = {2013},
}

@Article{AstropyII,
  author        = {{Astropy Collaboration} and {Price-Whelan}, A.~M. and {Sip{\H{o}}cz}, B.~M. and {G{\"u}nther}, H.~M. and {Lim}, P.~L. and {Crawford}, S.~M. and {Conseil}, S. and {Shupe}, D.~L. and {Craig}, M.~W. and {Dencheva}, N. and {Ginsburg}, A. and {VanderPlas}, J.~T. and {Bradley}, L.~D. and {P{\'e}rez-Su{\'a}rez}, D. and {de Val-Borro}, M. and {Aldcroft}, T.~L. and {Cruz}, K.~L. and {Robitaille}, T.~P. and {Tollerud}, E.~J. and {Ardelean}, C. and {Babej}, T. and {Bach}, Y.~P. and {Bachetti}, M. and {Bakanov}, A.~V. and {Bamford}, S.~P. and {Barentsen}, G. and {Barmby}, P. and {Baumbach}, A. and {Berry}, K.~L. and {Biscani}, F. and {Boquien}, M. and {Bostroem}, K.~A. and {Bouma}, L.~G. and {Brammer}, G.~B. and {Bray}, E.~M. and {Breytenbach}, H. and {Buddelmeijer}, H. and {Burke}, D.~J. and {Calderone}, G. and {Cano Rodr{\'\i}guez}, J.~L. and {Cara}, M. and {Cardoso}, J.~V.~M. and {Cheedella}, S. and {Copin}, Y. and {Corrales}, L. and {Crichton}, D. and {D'Avella}, D. and {Deil}, C. and {Depagne}, {\'E}. and {Dietrich}, J.~P. and {Donath}, A. and {Droettboom}, M. and {Earl}, N. and {Erben}, T. and {Fabbro}, S. and {Ferreira}, L.~A. and {Finethy}, T. and {Fox}, R.~T. and {Garrison}, L.~H. and {Gibbons}, S.~L.~J. and {Goldstein}, D.~A. and {Gommers}, R. and {Greco}, J.~P. and {Greenfield}, P. and {Groener}, A.~M. and {Grollier}, F. and {Hagen}, A. and {Hirst}, P. and {Homeier}, D. and {Horton}, A.~J. and {Hosseinzadeh}, G. and {Hu}, L. and {Hunkeler}, J.~S. and {Ivezi{\'c}}, {\v{Z}}. and {Jain}, A. and {Jenness}, T. and {Kanarek}, G. and {Kendrew}, S. and {Kern}, N.~S. and {Kerzendorf}, W.~E. and {Khvalko}, A. and {King}, J. and {Kirkby}, D. and {Kulkarni}, A.~M. and {Kumar}, A. and {Lee}, A. and {Lenz}, D. and {Littlefair}, S.~P. and {Ma}, Z. and {Macleod}, D.~M. and {Mastropietro}, M. and {McCully}, C. and {Montagnac}, S. and {Morris}, B.~M. and {Mueller}, M. and {Mumford}, S.~J. and {Muna}, D. and {Murphy}, N.~A. and {Nelson}, S. and {Nguyen}, G.~H. and {Ninan}, J.~P. and {N{\"o}the}, M. and {Ogaz}, S. and {Oh}, S. and {Parejko}, J.~K. and {Parley}, N. and {Pascual}, S. and {Patil}, R. and {Patil}, A.~A. and {Plunkett}, A.~L. and {Prochaska}, J.~X. and {Rastogi}, T. and {Reddy Janga}, V. and {Sabater}, J. and {Sakurikar}, P. and {Seifert}, M. and {Sherbert}, L.~E. and {Sherwood-Taylor}, H. and {Shih}, A.~Y. and {Sick}, J. and {Silbiger}, M.~T. and {Singanamalla}, S. and {Singer}, L.~P. and {Sladen}, P.~H. and {Sooley}, K.~A. and {Sornarajah}, S. and {Streicher}, O. and {Teuben}, P. and {Thomas}, S.~W. and {Tremblay}, G.~R. and {Turner}, J.~E.~H. and {Terr{\'o}n}, V. and {van Kerkwijk}, M.~H. and {de la Vega}, A. and {Watkins}, L.~L. and {Weaver}, B.~A. and {Whitmore}, J.~B. and {Woillez}, J. and {Zabalza}, V. and {Astropy Contributors}},
  title         = {{The Astropy Project: Building an Open-science Project and Status of the v2.0 Core Package}},
  doi           = {10.3847/1538-3881/aabc4f},
  eid           = {123},
  eprint        = {1801.02634},
  number        = {3},
  pages         = {123},
  volume        = {156},
  adsurl        = {https://ui.adsabs.harvard.edu/abs/2018AJ....156..123A},
  archiveprefix = {arXiv},
  creationdate  = {2023-01-12T17:15:53},
  journal       = {\aj},
  month         = sep,
  primaryclass  = {astro-ph.IM},
  year          = {2018},
}

@Article{AstropyIII,
  author        = {{Astropy Collaboration} and {Price-Whelan}, Adrian M. and {Lim}, Pey Lian and {Earl}, Nicholas and {Starkman}, Nathaniel and {Bradley}, Larry and {Shupe}, David L. and {Patil}, Aarya A. and {Corrales}, Lia and {Brasseur}, C.~E. and {N{\"o}the}, Maximilian and {Donath}, Axel and {Tollerud}, Erik and {Morris}, Brett M. and {Ginsburg}, Adam and {Vaher}, Eero and {Weaver}, Benjamin A. and {Tocknell}, James and {Jamieson}, William and {van Kerkwijk}, Marten H. and {Robitaille}, Thomas P. and {Merry}, Bruce and {Bachetti}, Matteo and {G{\"u}nther}, H. Moritz and {Aldcroft}, Thomas L. and {Alvarado-Montes}, Jaime A. and {Archibald}, Anne M. and {B{\'o}di}, Attila and {Bapat}, Shreyas and {Barentsen}, Geert and {Baz{\'a}n}, Juanjo and {Biswas}, Manish and {Boquien}, M{\'e}d{\'e}ric and {Burke}, D.~J. and {Cara}, Daria and {Cara}, Mihai and {Conroy}, Kyle E. and {Conseil}, Simon and {Craig}, Matthew W. and {Cross}, Robert M. and {Cruz}, Kelle L. and {D'Eugenio}, Francesco and {Dencheva}, Nadia and {Devillepoix}, Hadrien A.~R. and {Dietrich}, J{\"o}rg P. and {Eigenbrot}, Arthur Davis and {Erben}, Thomas and {Ferreira}, Leonardo and {Foreman-Mackey}, Daniel and {Fox}, Ryan and {Freij}, Nabil and {Garg}, Suyog and {Geda}, Robel and {Glattly}, Lauren and {Gondhalekar}, Yash and {Gordon}, Karl D. and {Grant}, David and {Greenfield}, Perry and {Groener}, Austen M. and {Guest}, Steve and {Gurovich}, Sebastian and {Handberg}, Rasmus and {Hart}, Akeem and {Hatfield-Dodds}, Zac and {Homeier}, Derek and {Hosseinzadeh}, Griffin and {Jenness}, Tim and {Jones}, Craig K. and {Joseph}, Prajwel and {Kalmbach}, J. Bryce and {Karamehmetoglu}, Emir and {Ka{\l}uszy{\'n}ski}, Miko{\l}aj and {Kelley}, Michael S.~P. and {Kern}, Nicholas and {Kerzendorf}, Wolfgang E. and {Koch}, Eric W. and {Kulumani}, Shankar and {Lee}, Antony and {Ly}, Chun and {Ma}, Zhiyuan and {MacBride}, Conor and {Maljaars}, Jakob M. and {Muna}, Demitri and {Murphy}, N.~A. and {Norman}, Henrik and {O'Steen}, Richard and {Oman}, Kyle A. and {Pacifici}, Camilla and {Pascual}, Sergio and {Pascual-Granado}, J. and {Patil}, Rohit R. and {Perren}, Gabriel I. and {Pickering}, Timothy E. and {Rastogi}, Tanuj and {Roulston}, Benjamin R. and {Ryan}, Daniel F. and {Rykoff}, Eli S. and {Sabater}, Jose and {Sakurikar}, Parikshit and {Salgado}, Jes{\'u}s and {Sanghi}, Aniket and {Saunders}, Nicholas and {Savchenko}, Volodymyr and {Schwardt}, Ludwig and {Seifert-Eckert}, Michael and {Shih}, Albert Y. and {Jain}, Anany Shrey and {Shukla}, Gyanendra and {Sick}, Jonathan and {Simpson}, Chris and {Singanamalla}, Sudheesh and {Singer}, Leo P. and {Singhal}, Jaladh and {Sinha}, Manodeep and {Sip{\H{o}}cz}, Brigitta M. and {Spitler}, Lee R. and {Stansby}, David and {Streicher}, Ole and {{\v{S}}umak}, Jani and {Swinbank}, John D. and {Taranu}, Dan S. and {Tewary}, Nikita and {Tremblay}, Grant R. and {de Val-Borro}, Miguel and {Van Kooten}, Samuel J. and {Vasovi{\'c}}, Zlatan and {Verma}, Shresth and {de Miranda Cardoso}, Jos{\'e} Vin{\'\i}cius and {Williams}, Peter K.~G. and {Wilson}, Tom J. and {Winkel}, Benjamin and {Wood-Vasey}, W.~M. and {Xue}, Rui and {Yoachim}, Peter and {Zhang}, Chen and {Zonca}, Andrea and {Astropy Project Contributors}},
  title         = {{The Astropy Project: Sustaining and Growing a Community-oriented Open-source Project and the Latest Major Release (v5.0) of the Core Package}},
  doi           = {10.3847/1538-4357/ac7c74},
  eid           = {167},
  eprint        = {2206.14220},
  number        = {2},
  pages         = {167},
  volume        = {935},
  adsurl        = {https://ui.adsabs.harvard.edu/abs/2022ApJ...935..167A},
  archiveprefix = {arXiv},
  journal       = {\apj},
  month         = aug,
  primaryclass  = {astro-ph.IM},
  year          = {2022},
}

@Article{hunter2007matplotlib,
  author        = {Hunter, John D},
  title         = {Matplotlib: A 2D graphics environment},
  number        = {3},
  pages         = {90},
  volume        = {9},
  journal       = {Computing in science \& engineering},
  publisher     = {IEEE Computer Society},
  year          = {2007},
}

@InProceedings{McKinney_2010,
  author       = {{W}es {M}c{K}inney},
  booktitle    = {{P}roceedings of the 9th {P}ython in {S}cience {C}onference},
  title        = {{D}ata {S}tructures for {S}tatistical {C}omputing in {P}ython},
  doi          = {10.25080/Majora-92bf1922-00a},
  editor       = {{S}t\'efan van der {W}alt and {J}arrod {M}illman},
  pages        = {56 - 61},
  creationdate = {2023-01-12T17:11:54},
  year         = {2010},
}

@Software{reback2020pandas,
  author       = {The pandas development team},
  title        = {pandas-dev/pandas: Pandas},
  doi          = {10.5281/zenodo.3509134},
  url          = {https://doi.org/10.5281/zenodo.3509134},
  version      = {1.5.0},
  creationdate = {2023-01-12T17:13:27},
  month        = feb,
  publisher    = {Zenodo},
  year         = {2020},
}

@Misc{jones2001scipy,
  author        = {Jones, Eric and Oliphant, Travis and Peterson, Pearu and others},
  title         = {{SciPy}: Open source scientific tools for {Python}},
  url           = {http://www.scipy.org/},
  description   = {Citing SciPy library --- SciPy.org},
  year          = {2024},
}

@Article{TIC2019,
  author        = {{Stassun}, Keivan G. and {Oelkers}, Ryan J. and {Paegert}, Martin and {Torres}, Guillermo and {Pepper}, Joshua and {De Lee}, Nathan and {Collins}, Kevin and {Latham}, David W. and {Muirhead}, Philip S. and {Chittidi}, Jay and {Rojas-Ayala}, B{\'a}rbara and {Fleming}, Scott W. and {Rose}, Mark E. and {Tenenbaum}, Peter and {Ting}, Eric B. and {Kane}, Stephen R. and {Barclay}, Thomas and {Bean}, Jacob L. and {Brassuer}, C.~E. and {Charbonneau}, David and {Ge}, Jian and {Lissauer}, Jack J. and {Mann}, Andrew W. and {McLean}, Brian and {Mullally}, Susan and {Narita}, Norio and {Plavchan}, Peter and {Ricker}, George R. and {Sasselov}, Dimitar and {Seager}, S. and {Sharma}, Sanjib and {Shiao}, Bernie and {Sozzetti}, Alessandro and {Stello}, Dennis and {Vanderspek}, Roland and {Wallace}, Geoff and {Winn}, Joshua N.},
  title         = {{The Revised TESS Input Catalog and Candidate Target List}},
  doi           = {10.3847/1538-3881/ab3467},
  eid           = {138},
  eprint        = {1905.10694},
  number        = {4},
  pages         = {138},
  volume        = {158},
  adsurl        = {https://ui.adsabs.harvard.edu/abs/2019AJ....158..138S},
  archiveprefix = {arXiv},
  journal       = {\aj},
  month         = oct,
  primaryclass  = {astro-ph.SR},
  year          = {2019},
}

@ARTICLE{pax11,
       author = {{Paxton}, Bill and {Bildsten}, Lars and {Dotter}, Aaron and
         {Herwig}, Falk and {Lesaffre}, Pierre and {Timmes}, Frank},
        title = "{Modules for Experiments in Stellar Astrophysics (MESA)}",
      journal = {\apjs},
         year = 2011,
        month = jan,
       volume = {192},
       number = {1},
          eid = {3},
        pages = {3},
          doi = {10.1088/0067-0049/192/1/3},
archivePrefix = {arXiv},
       eprint = {1009.1622},
 primaryClass = {astro-ph.SR},
       adsurl = {https://ui.adsabs.harvard.edu/abs/2011ApJS..192....3P}
}

@ARTICLE{pax13,
       author = {{Paxton}, Bill and {Cantiello}, Matteo and {Arras}, Phil and
         {Bildsten}, Lars and {Brown}, Edward F. and {Dotter}, Aaron and
         {Mankovich}, Christopher and {Montgomery}, M.~H. and {Stello}, Dennis and
         {Timmes}, F.~X. and {Townsend}, Richard},
        title = "{Modules for Experiments in Stellar Astrophysics (MESA): Planets, Oscillations, Rotation, and Massive Stars}",
      journal = {\apjs},
         year = 2013,
        month = sep,
       volume = {208},
       number = {1},
          eid = {4},
        pages = {4},
          doi = {10.1088/0067-0049/208/1/4},
archivePrefix = {arXiv},
       eprint = {1301.0319},
 primaryClass = {astro-ph.SR},
       adsurl = {https://ui.adsabs.harvard.edu/abs/2013ApJS..208....4P}
}

@ARTICLE{pax18,
       author = {{Paxton}, Bill and {Schwab}, Josiah and {Bauer}, Evan B. and
         {Bildsten}, Lars and {Blinnikov}, Sergei and {Duffell}, Paul and
         {Farmer}, R. and {Goldberg}, Jared A. and {Marchant}, Pablo and
         {Sorokina}, Elena and {Thoul}, Anne and {Townsend}, Richard H.~D. and
         {Timmes}, F.~X.},
        title = "{Modules for Experiments in Stellar Astrophysics (MESA): Convective Boundaries, Element Diffusion, and Massive Star Explosions}",
      journal = {\apjs},
         year = 2018,
        month = feb,
       volume = {234},
       number = {2},
          eid = {34},
        pages = {34},
          doi = {10.3847/1538-4365/aaa5a8},
archivePrefix = {arXiv},
       eprint = {1710.08424},
 primaryClass = {astro-ph.SR},
       adsurl = {https://ui.adsabs.harvard.edu/abs/2018ApJS..234...34P}
}

@ARTICLE{pax19,
       author = {{Paxton}, Bill and {Smolec}, R. and {Schwab}, Josiah and {Gautschy}, A. and {Bildsten}, Lars and {Cantiello}, Matteo and {Dotter}, Aaron and {Farmer}, R. and {Goldberg}, Jared A. and {Jermyn}, Adam S. and {Kanbur}, S.~M. and {Marchant}, Pablo and {Thoul}, Anne and {Townsend}, Richard H.~D. and {Wolf}, William M. and {Zhang}, Michael and {Timmes}, F.~X.},
        title = "{Modules for Experiments in Stellar Astrophysics (MESA): Pulsating Variable Stars, Rotation, Convective Boundaries, and Energy Conservation}",
      journal = {\apjs},
         year = 2019,
        month = jul,
       volume = {243},
       number = {1},
          eid = {10},
        pages = {10},
          doi = {10.3847/1538-4365/ab2241},
archivePrefix = {arXiv},
       eprint = {1903.01426},
 primaryClass = {astro-ph.SR},
       adsurl = {https://ui.adsabs.harvard.edu/abs/2019ApJS..243...10P}
}

@ARTICLE{jer23,
       author = {{Jermyn}, Adam S. and {Bauer}, Evan B. and {Schwab}, Josiah and {Farmer}, R. and {Ball}, Warrick H. and {Bellinger}, Earl P. and {Dotter}, Aaron and {Joyce}, Meridith and {Marchant}, Pablo and {Mombarg}, Joey S.~G. and {Wolf}, William M. and {Sunny Wong}, Tin Long and {Cinquegrana}, Giulia C. and {Farrell}, Eoin and {Smolec}, R. and {Thoul}, Anne and {Cantiello}, Matteo and {Herwig}, Falk and {Toloza}, Odette and {Bildsten}, Lars and {Townsend}, Richard H.~D. and {Timmes}, F.~X.},
        title = "{Modules for Experiments in Stellar Astrophysics (MESA): Time-dependent Convection, Energy Conservation, Automatic Differentiation, and Infrastructure}",
      journal = {\apjs},
         year = 2023,
        month = mar,
       volume = {265},
       number = {1},
          eid = {15},
        pages = {15},
          doi = {10.3847/1538-4365/acae8d},
archivePrefix = {arXiv},
       eprint = {2208.03651},
 primaryClass = {astro-ph.SR},
       adsurl = {https://ui.adsabs.harvard.edu/abs/2023ApJS..265...15J}
}

@ARTICLE{cho16,
       author = {{Choi}, Jieun and {Dotter}, Aaron and {Conroy}, Charlie and
         {Cantiello}, Matteo and {Paxton}, Bill and {Johnson}, Benjamin D.},
        title = "{Mesa Isochrones and Stellar Tracks (MIST). I. Solar-scaled Models}",
      journal = {\apj},
         year = 2016,
        month = jun,
       volume = {823},
       number = {2},
          eid = {102},
        pages = {102},
          doi = {10.3847/0004-637X/823/2/102},
archivePrefix = {arXiv},
       eprint = {1604.08592},
 primaryClass = {astro-ph.SR},
       adsurl = {https://ui.adsabs.harvard.edu/abs/2016ApJ...823..102C}
}

@ARTICLE{dot16,
       author = {{Dotter}, Aaron},
        title = "{MESA Isochrones and Stellar Tracks (MIST) 0: Methods for the Construction of Stellar Isochrones}",
      journal = {\apjs},
         year = 2016,
        month = jan,
       volume = {222},
       number = {1},
          eid = {8},
        pages = {8},
          doi = {10.3847/0067-0049/222/1/8},
archivePrefix = {arXiv},
       eprint = {1601.05144},
 primaryClass = {astro-ph.SR},
       adsurl = {https://ui.adsabs.harvard.edu/abs/2016ApJS..222....8D}
}

@article{gala,
  doi = {10.21105/joss.00388},
  url = {https://doi.org/10.21105%2Fjoss.00388},
  year = 2017,
  month = {oct},
  publisher = {The Open Journal},
  volume = {2},
  number = {18},
  author = {Adrian M. Price-Whelan},
  title = {Gala: A Python package for galactic dynamics},
  journal = {The Journal of Open Source Software}}

@software{adrian_price_whelan_2020_4159870,
  author       = {Adrian Price-Whelan and
                  Brigitta Sipőcz and
                  Daniel Lenz and
                  Johnny Greco and
                  Nathaniel Starkman and
                  Dan Foreman-Mackey and
                  P. L. Lim and
                  Semyeong Oh and
                  Sergey Koposov and
                  Syrtis Major},
  title        = {adrn/gala: v1.3},
  month        = oct,
  year         = 2020,
  publisher    = {Zenodo},
  version      = {v1.8.1},
  doi          = {10.5281/zenodo.4159870},
  url          = {https://doi.org/10.5281/zenodo.4159870},
}

@Article{harris2020array,
  author        = {Charles R. Harris and K. Jarrod Millman and St{\'{e}}fan J. van der Walt and Ralf Gommers and Pauli Virtanen and David Cournapeau and Eric Wieser and Julian Taylor and Sebastian Berg and Nathaniel J. Smith and Robert Kern and Matti Picus and Stephan Hoyer and Marten H. van Kerkwijk and Matthew Brett and Allan Haldane and Jaime Fern{\'{a}}ndez del R{\'{i}}o and Mark Wiebe and Pearu Peterson and Pierre G{\'{e}}rard-Marchant and Kevin Sheppard and Tyler Reddy and Warren Weckesser and Hameer Abbasi and Christoph Gohlke and Travis E. Oliphant},
  title         = {Array programming with {NumPy}},
  doi           = {10.1038/s41586-020-2649-2},
  number        = {7825},
  pages         = {357--362},
  url           = {https://doi.org/10.1038/s41586-020-2649-2},
  volume        = {585},
  journal       = {Nature},
  month         = sep,
  publisher     = {Springer Science and Business Media {LLC}},
  year          = {2020},
}

@Software{adrian_price_whelan_2018_1228136,
  author       = {Adrian Price-Whelan},
  title        = {adrn/pyia: v0.2},
  doi          = {10.5281/zenodo.1228136},
  url          = {https://doi.org/10.5281/zenodo.1228136},
  version      = {v0.2},
  creationdate = {2023-03-07T13:17:49},
  month        = apr,
  publisher    = {Zenodo},
  year         = {2018},
}

@Article{2020SciPy-NMeth,
  author        = {Virtanen, Pauli and Gommers, Ralf and Oliphant, Travis E. and Haberland, Matt and Reddy, Tyler and Cournapeau, David and Burovski, Evgeni and Peterson, Pearu and Weckesser, Warren and Bright, Jonathan and {van der Walt}, St{\'e}fan J. and Brett, Matthew and Wilson, Joshua and Millman, K. Jarrod and Mayorov, Nikolay and Nelson, Andrew R. J. and Jones, Eric and Kern, Robert and Larson, Eric and Carey, C J and Polat, {\.I}lhan and Feng, Yu and Moore, Eric W. and {VanderPlas}, Jake and Laxalde, Denis and Perktold, Josef and Cimrman, Robert and Henriksen, Ian and Quintero, E. A. and Harris, Charles R. and Archibald, Anne M. and Ribeiro, Ant{\^o}nio H. and Pedregosa, Fabian and {van Mulbregt}, Paul and {SciPy 1.0 Contributors}},
  title         = {{{SciPy} 1.0: Fundamental Algorithms for Scientific Computing in Python}},
  doi           = {10.1038/s41592-019-0686-2},
  pages         = {261--272},
  volume        = {17},
  adsurl        = {https://rdcu.be/b08Wh},
  journal       = {Nature Methods},
  year          = {2020},
}

@Article{Leconte10,
  author        = {{Leconte}, J. and {Chabrier}, G. and {Baraffe}, I. and {Levrard}, B.},
  title         = {{Is tidal heating sufficient to explain bloated exoplanets? Consistent calculations accounting for finite initial eccentricity}},
  doi           = {10.1051/0004-6361/201014337},
  eid           = {A64},
  eprint        = {1004.0463},
  pages         = {A64},
  volume        = {516},
  adsurl        = {https://ui.adsabs.harvard.edu/abs/2010A&A...516A..64L},
  archiveprefix = {arXiv},
  journal       = {\aap},
  month         = jun,
  primaryclass  = {astro-ph.EP},
  year          = {2010},
}
\bibliographystyle{aasjournal}

\end{document}